\documentclass[12pt]{article}
\pdfoutput=1
\usepackage[nosort]{cite}

\usepackage[svgnames,psnames]{xcolor}
\usepackage{draft,hyperref,tikz,float,subfig,cite}
 
\usepackage{tikz-cd}
\usetikzlibrary{calc,topaths,decorations,decorations.pathmorphing,arrows,decorations.markings,cd}

\usetikzlibrary{calc,shapes.geometric}
\pgfmathparse{atan2(0,1)}
\ifdim\pgfmathresult pt=0pt
  \tikzset{declare function={atanXY(\x,\y)=atan2(\y,\x);atanYX(\y,\x)=atan2(\y,\x);}}
\else                 
  \tikzset{declare function={atanXY(\x,\y)=atan2(\x,\y);atanYX(\y,\x)=atan2(\x,\y);}}
\fi

\usetikzlibrary{snakes}
\usetikzlibrary{shapes.misc}
\usetikzlibrary{decorations.markings}

\tikzset{line/.style={line width=0.25mm},
curve/.style={line,smooth,tension=1},
->-/.style={decoration={
  markings,
  mark=at position #1 with {\arrow[>=stealth]{>}}},postaction={decorate}},
-<-/.style={decoration={
  markings,
  mark=at position #1 with {\arrow[>=stealth]{<}}},postaction={decorate}},
}

\newcommand{\cH}{\mathcal{H}}
\newcommand{\cZ}{\mathcal{Z}}
\newcommand{\ZC}{\mathcal{Z(C)}}

\newcommand{\cO}{\mathcal{O}}
\newcommand{\TV}{\mathrm{TV}}

\renewcommand{\S}{S}
\renewcommand{\D}{\mathbb{D}}
\renewcommand{\B}{\mathbb{B}}
\newcommand{\T}{T}
\newcommand{\Cen}{\mathrm{C}}

\newcommand{\bB}{\mathbb{B}}

\newcommand{\SL}{\mathrm{SL}}

\begin{document}

\begin{titlepage}

\begin{center}

\hfill YITP-SB-2023-03
\\
\vfill
\title{Bootstrapping Non-Invertible Symmetries}

\author{Ying-Hsuan Lin$^{a}$ and Shu-Heng Shao$^b$
}

\address{${}^a$Jefferson Physical Laboratory, Harvard University,  Cambridge, MA 02138, USA}
\address{${}^b$C.\ N.\ Yang Institute for Theoretical Physics, Stony Brook University, Stony Brook, NY 11794, USA}

\email{yhlin@fas.harvard.edu, shu-heng.shao@stonybrook.edu}

\end{center}

\abstract{
Using the numerical modular bootstrap, we constrain the space of 1+1d CFTs with a finite non-invertible global symmetry described by a fusion category $\mathcal{C}$. 
We derive universal and rigorous upper bounds on the lightest $\mathcal{C}$-preserving scalar local operator for fusion categories such as the Ising and Fibonacci categories. 
These numerical bounds constrain the possible robust gapless phases protected by a non-invertible global symmetry, which commonly arise from microscopic lattice models such as the anyonic chains. 
We also derive bounds on the lightest $\mathcal{C}$-violating local operator.
Our bootstrap equations naturally arise from a ``slab construction", where the CFT is coupled to the 2+1d Turaev-Viro TQFT, also known as the Symmetry TFT.
}

\vfill

\end{titlepage}

\eject

\tableofcontents

\section{Introduction}

In recent years, we have seen rapid developments of generalized global symmetries \cite{Gaiotto:2014kfa} in high energy physics, condensed matter physics, and mathematics. 
See \cite{McGreevy:2022oyu,Cordova:2022ruw} for reviews. 
Among these exciting developments is the study of an exotic kind of symmetry that does not form a group. In particular, not every symmetry transformation is invertible. 
These \textit{non-invertible symmetries} have been most systematically analyzed and best understood in the context of 1+1d conformal field theories (CFT) through a series of developments \cite{Verlinde:1988sn,Petkova:2000ip,Fuchs:2002cm,Frohlich:2004ef,Frohlich:2006ch,Feiguin:2006ydp,Fuchs:2007tx,Fredenhagen:2009tn,Frohlich:2009gb,Carqueville:2012dk,Aasen:2016dop,Bhardwaj:2017xup,Tachikawa:2017gyf,Chang:2018iay,Ji:2019ugf,Cordova:2019wpi,Lin:2019hks,Thorngren:2019iar,Huang:2020lox,Gaiotto:2020iye,Komargodski:2020mxz,Aasen:2020jwb,Chang:2020imq,Huang:2021ytb,Thorngren:2021yso,Huang:2021zvu,Huang:2021nvb,Vanhove:2021zop,Burbano:2021loy,Inamura:2022lun,Chang:2022hud,Lin:2022dhv}, building on the seminal work of \cite{Moore:1988qv,Moore:1989yh}. 
They have also been generalized to a variety of higher-dimensional quantum systems, including the real-world QED \cite{Choi:2022jqy,Cordova:2022ieu,Choi:2022rfe}.
See \cite{Rudelius:2020orz,Heidenreich:2021xpr,Nguyen:2021yld,Wang:2021vki,Koide:2021zxj,Choi:2021kmx,Kaidi:2021xfk,Roumpedakis:2022aik,Bhardwaj:2022yxj,Choi:2022zal,Bhardwaj:2022lsg,Bartsch:2022mpm,Apruzzi:2022rei,GarciaEtxebarria:2022vzq,Heckman:2022muc,Freed:2022qnc,Niro:2022ctq,Antinucci:2022vyk,Chen:2022cyw,Cordova:2022fhg,Karasik:2022kkq,GarciaEtxebarria:2022jky,Choi:2022fgx,Yokokura:2022alv,Bhardwaj:2022kot,Bartsch:2022ytj,Bhardwaj:2022maz} for a partial list of recent advancements in non-invertible symmetries in general spacetime dimensions.

 In 1+1d relativistic quantum field theory, generalized global symmetries are implemented by topological line operators/defects in Euclidean spacetime.  
 The mathematical framework for a finite (invertible or non-invertible) generalized global symmetry is a \textit{fusion category} \cite{etingof2016tensor}. 
When the fusion of some of these lines is not group-like, they generate a non-invertible global symmetry. 
Non-invertible symmetries are ubiquitous in 1+1d CFT, with the simplest example being the Kramers-Wannier duality line in the critical Ising CFT \cite{Frohlich:2004ef}. 
These generalized symmetries have dramatic consequences on quantum systems and have led to nontrivial constraints on renormalization group flows \cite{Chang:2018iay,Thorngren:2019iar,Komargodski:2020mxz,Thorngren:2021yso} and selection rules \cite{Lin:2022dhv}.

New symmetries give rise to new notions of naturalness \cite{tHooft:1979rat}.  
Starting from a microscopic system with a generalized global symmetry $\cC$,  it is natural to impose this symmetry to forbid any $\cC$-violating relevant deformation of the low-energy phase.  
For instance, there is a large class of 1+1d anyonic chains \cite{Feiguin:2006ydp,Gils:2013swa,Buican:2017rxc,Huang:2021nvb,Vanhove:2021zop,Liu:2022qwn} that realize non-invertible symmetries. 
See also \cite{Hauru:2015abi,Aasen:2016dop,Aasen:2020jwb,Choi:2021kmx,Hayashi:2022fkw,Apte:2022xtu,Delcamp:2023kew} for statistical models and other lattice models with non-invertible symmetries. 
The low-energy theories are often robust (or stable) gapless phases protected by a generalized global symmetry. 
For example, the famous ``golden chain" \cite{Feiguin:2006ydp} flows to the tricritical Ising CFT,  which would not have been robust and numerically realizable had it been only protected by its conventional group-like symmetry, $\bZ_2$. 
(See \cite{Seiberg:2020bhn} for a recent discussion on robustness and naturalness in quantum field theory.)

What are the universal properties of a general 1+1d CFT with a finite non-invertible symmetry $\cC$ for such a gapless phase? 
We approach this question with one of the most effective non-perturbative tools to study CFT, the conformal bootstrap \cite{Rattazzi:2008pe,Poland:2018epd}.  
More precisely, we will constrain the space of 1+1d CFT with a non-invertible symmetry from modular covariance.

Modular consistency is the condition that the partition function $Z_{\cM_{d+1}}$ of a general $(d+1)$-dimensional quantum field theory on an arbitrary spacetime manifold $\cM_{d+1}$ must be invariant under the mapping class group of $\cM_{d+1}$.  When $\cM_{d+1} = \S^1 \times M_d$, this partition function can be computed as a trace over the Hilbert space $\cH_{M_d}$ of the theory quantized on $M_d$.  If furthermore $\cM_{d+1} = \S^1_a \times \S^1_b \times \mu_{d-1}$, then the fact that $Z_{\cM_{d+1}}$ can be evaluated as traces over $\cH_{\S^1_a \times \mu_{d-1}}$ and over $\cH_{\S^1_b \times \mu_{d-1}}$ gives a relation between the two Hilbert spaces. 
Modular bootstrap is the idea that for a (1+1)-dimensional CFT, the Hamiltonians on $\cH_{\S^1_a}$ and on $\cH_{\S^1_b}$ are related by a simple rescaling, and hence the invariance of $Z_{\T^2}$ under the mapping class group $\SL(2,\bZ)$ imposes strong constraints on the Hilbert space on a circle. 
Cardy applied this idea to produce his seminal formula \cite{Cardy:1986ie}.
The modern numerical modular bootstrap was initiated in \cite{Hellerman:2009bu} and has since been generalized in many different ways in 1+1d \cite{Hellerman:2010qd,Keller:2012mr,Friedan:2013cba,Qualls:2013eha,Hartman:2014oaa,Qualls:2014oea,Kim:2015oca,Benjamin:2016fhe,Montero:2016tif,Collier:2016cls,Collier:2017shs,Cho:2017fzo,Keller:2017iql,Cardy:2017qhl,Bae:2017kcl,Dyer:2017rul,Anous:2018hjh,Bae:2018qym,Afkhami-Jeddi:2019zci,Mukhametzhanov:2019pzy,Lin:2019kpn,Hartman:2019pcd,Ganguly:2019ksp,Benjamin:2019stq,Pal:2019zzr,Alday:2019vdr,Benjamin:2020swg,Mukhametzhanov:2020swe,Pal:2020wwd,Benjamin:2020zbs,Afkhami-Jeddi:2020hde,Dymarsky:2020bps,Dymarsky:2020qom,Alday:2020qkm,Benjamin:2021ygh,Collier:2021ngi,Benjamin:2022pnx}.\footnote{Modular consistency outside the 1+1d CFT context has also been effectively utilized in, for example, \cite{Shaghoulian:2015kta,Belin:2016yll,Luo:2022tqy}.}

In the presence of a global symmetry, one can further refine the partition function by insertions of the topological lines. 
These twisted partition functions obey certain modular covariant conditions, depending on the global symmetry and its 't Hooft anomaly. 
The numerical modular bootstrap has been applied to obtain universal constraints on CFT with invertible global symmetries  \cite{Benjamin:2016fhe,Montero:2016tif,Lin:2019kpn,Pal:2020wwd,Lin:2021udi,Grigoletto:2021zyv,Cao:2021euf,Lanzetta:2022lze}. 
The implications of the 't Hooft anomalies on the bootstrap bounds were first studied in \cite{Lin:2019kpn} and further generalized in \cite{Lin:2021udi,Grigoletto:2021zyv,Lanzetta:2022lze}.

In this paper, we apply the numerical modular bootstrap to constrain 1+1d CFT with a finite non-invertible global symmetry. 
We focus on the three simplest non-invertible symmetries, the Fibonacci, Ising, and $\mathfrak{su}(2)_2$ fusion categories.  
We derive rigorous and universal bounds on the local and non-local operators in any CFT with the above non-invertible symmetry.

Our modular bootstrap equations can be compactly summarized using a ``slab construction" by coupling the CFT to a TQFT in one dimension higher. 
This slab construction has recently received a lot of attention in the study of generalized symmetries \cite{Gaiotto:2020iye,Freed:2022qnc,Lin:2022dhv}, which sometimes goes under the name of Symmetry TFT \cite{2015arXiv150201690K,Apruzzi:2021nmk,Freed:2022qnc,Moradi:2022lqp,Kaidi:2022cpf,Kaidi:2023maf}, and also in the context of categorical symmetries \cite{Ji:2019eqo,Ji:2019jhk,Ji:2021esj,Chatterjee:2022tyg,Chatterjee:2022jll}. 
More specifically, we couple the CFT to a 2+1d Turaev-Viro TQFT $\mathcal{Z}(\cC)$ by gauging the non-invertible symmetry $\cC$. 
The partition functions $Z^\text{3d}_\mu$ of this 2+1d system, each labeled by an anyon line $\mu$, transform according to the $S$ and $T$ matrices of the Turaev-Viro TQFT. 
Applying the numerical modular bootstrap to this 2+1d system yields universal bounds on the states in different anyon sectors of this 2+1d system. 
Finally, by relating the Hilbert spaces in 1+1d and 2+1d (see \eqref{HV}), we  translate these bounds into universal constraints on the 1+1d CFT. 
We emphasize that we use the 2+1d TQFT only as a trick in the intermediate steps, while the final numerical bounds apply to any 1+1d CFT with a non-invertible symmetry without a bulk.

In particular, we derive upper bounds on the lightest $\cC$-preserving scalar operator in any CFT.  
For each of the symmetries above, we find a window of the central charge such that the bound is below $\Delta=2$. 
Therefore, our results rigorously rule out stable CFT in those ranges of the central charge if we only impose the non-invertible symmetry $\cC$. 
This has potential implications on the phase diagram for microscopic lattice models with a non-invertible symmetry, such as the anyonic chains.

This paper is organized as follows. In Section \ref{sec:TDL} we briefly review non-invertible symmetries and their topological line defects in 1+1d CFT. 
In Section \ref{Sec:Examples}, we review three different non-invertible symmetries, the Fibonacci, Ising, and $\mathfrak{su}(2)_2$ fusion categories, that we will later bootstrap. 
In Section \ref{sec:anyon} we couple a general 1+1d CFT with a non-invertible symmetry $\cC$ to the 2+1d Turaev-Viro TQFT $\mathcal{Z}(\cC)$. We relate the twisted Hilbert spaces of the 1+1d CFT to Hilbert spaces of the 2+1d system with an anyon line insertion. The partition functions for the latter obey a simple modular property that we review. 
Section \ref{sec:modular} explains our bootstrap method and the numerical bootstrap bounds. 
Finally, Appendix \ref{app:existence} explains when a bootstrap bound exists in a given sector.

\section{Non-invertible Topological Lines in 1+1 Dimensions}\label{sec:TDL}
 
In this section, we discuss some general properties of topological lines in the context of unitary CFT.  
We will only consider bosonic theories which do not require a choice of the spin structure. 
Throughout the entire paper, we will assume that the CFT has a unique topological local operator, namely, the identity operator. 
We only consider CFT with equal left and right central charges, i.e.\ $c = c_L=c_R$. 
Our discussion below is by no means comprehensive, and we refer the reader to  \cite{Bhardwaj:2017xup,Chang:2018iay} and references therein for a more detailed discussion.

\bigskip\bigskip\centerline{\it Ordinary Global Symmetries and Invertible Lines}\bigskip

In relativistic systems, the modern formulation of global symmetries is in terms of their topological symmetry operators or defects \cite{Gaiotto:2014kfa}. 
In 1+1-dimensional CFT, an ordinary (zero-form), continuous or discrete, global symmetry $G$ is implemented by a codimension-one topological defect line in spacetime. 
Such a topological line ${\cal L}_g$ is labeled by a group element $g\in G$.  
The fusion rule of topological lines is defined by placing two parallel topological lines wrapped around a cylinder and bringing them close to each other. 
For topological lines associated with an ordinary global symmetry $G$, their fusion obeys the group multiplication law (which may be abelian or non-abelian): 
\ie
{\cal L}_{g_1} \times {\cal L}_{g_2} = {\cal L}_{g_1g_2}\,,~~~g_1,g_2\in G\,.
\fe
In particular, every such line ${\cal L}_g$ has an inverse ${\cal L}_{g^{-1}}$ such that the fusion between the two is  ${\cal L}_g\times {\cal L}_{g^{-1}} = {\cal L}_{g^{-1}}\times {\cal L}_g =1$, 
where 1 is the trivial identity line.  
For this reason, these lines associated with an ordinary global symmetry $G$ are called \textit{invertible}.

Given two lines, we can consider their direct sum, ${\cal L}\equiv {\cal L}_1+{\cal L}_2$.  
The correlation function of $\cal L$ is a sum of those of ${\cal L}_1$ and ${\cal L}_2$, i.e.\ $\langle \cdots{\cal L}\rangle=\langle \cdots{\cal L}_1\rangle+\langle \cdots{\cal L}_2\rangle$.  
More generally, one can define a linear combination of lines with non-negative integer coefficients. 
A line is called simple if it is not a linear combination of other lines. 

\bigskip\bigskip\centerline{\it Non-invertible Topological Lines}\bigskip

In almost every known 1+1d CFT, there are topological lines that are not associated with any ordinary global symmetry.  
The fusion rule of such simple lines takes a more general form:
\ie\label{fusion}
{\cal L}_a \times {\cal L}_b = \sum_c N^c_{ab} {\cal L}_c\,,~~~N^c_{ab}\in \mathbb{Z}_{\ge0}\,.
\fe
This fusion is not group-like because there is generally more than one term on the right-hand side. 
In particular, not every line has an inverse. 
A line without an inverse is called \textit{non-invertible} and is not associated with any ordinary global symmetry. 
In 1+1d unitary QFT, a finite set of topological lines form a   unitary fusion category $\cC$, generalizing the notion of a finite group.

For every simple (invertible or not) line ${\cal L}$, there is an orientation-reversed line $\overline {\cal L}$, known as the dual object in the fusion category, such that ${\cal L}\times \overline {\cal L} = \overline{\cal L}\times {\cal L}=1+\cdots$ contains the identity line. 
Finally, the coefficient $N^c_{ab} \in \mathbb{Z}_{\ge0}$ in the fusion rule \eqref{fusion} is a non-negative integer, which is the dimension of the Hilbert space of topological operators living at the junction between ${\cal L}_a, {\cal L}_b,\bar {\cal L}_c$.

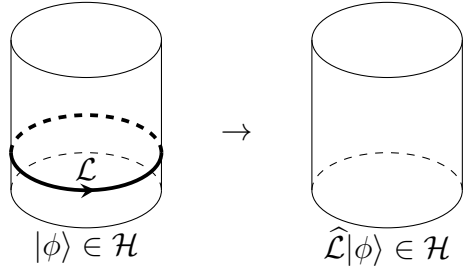
\begin{figure}
\centering
\begin{tikzpicture}[scale=.5]
\draw (0,0) ++ (180:2 and 1) arc (180:360:2 and 1);
\draw [dashed] (0,0) ++ (0:2 and 1) arc (0:180:2 and 1);
\draw (0,4) ellipse (2 and 1);
\draw (-2,0) -- (-2,4);
\draw (2,0) -- (2,4);
\draw [line,->-=0.55,line width=0.5mm] (0,1) ++ (180:2 and 1) arc (180:360:2 and 1);
\node at (0,.55) {$\cal L$} ;
\draw [line width=0.5mm,dashed] (0,1) ++ (0:2 and 1) arc (0:180:2 and 1);
\node at (0,-1.5) {$|\phi\rangle \in {\cal H}$};
\node at (4,1.5) {$\rightarrow$};
\begin{scope}[xshift = 8cm]
\draw (0,0) ++ (180:2 and 1) arc (180:360:2 and 1);
\draw [dashed] (0,0) ++ (0:2 and 1) arc (0:180:2 and 1);
\draw (0,4) ellipse (2 and 1);
\draw (-2,0) -- (-2,4);
\draw (2,0) -- (2,4);
\node at (0,-1.5) {$\widehat\cL |\phi\rangle \in {\cal H}$};
\end{scope}
\end{tikzpicture}
\caption{A topological defect line $\cal L$ wrapped around the spatial circle of a cylinder leads to an action $\widehat \cL$ on the Hilbert space $\cal H$.
}
\label{fig:L}
\end{figure}

Given a general topological line $\cal L$, which may or may not be invertible, we can define an action $\widehat{\cal L}:~{\cal H}\to {\cal H}$ on the Hilbert space $\cal H$ of local operators by wrapping the line around a cylinder as in Figure \ref{fig:L}. 
For an invertible line associated with a global symmetry $G$, this action gives the symmetry transformation of a local operator, i.e.\ $\widehat{\cal L}_g: \phi \to g\cdot \phi$. 
For a non-invertible line, the corresponding action $\widehat{\cal L}$ might not be invertible on some local operators. For example, it may map a nontrivial local operator to 0 (not to be confused with the identity operator).

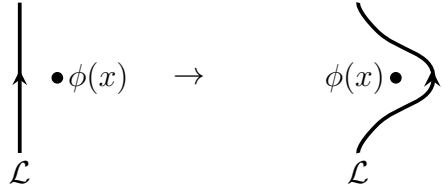
\begin{figure}
\centering
~~~~
\begin{tikzpicture}[scale=.5]
\draw [line,->-=0.55,line width=0.5mm] (0,0) -- (0,4);
\node at (0,-.5) {$\cal L$};
\filldraw (1,2) circle (4pt) node[right=0mm] {$\phi(x)$};
\node at (4.5,2) {$\rightarrow$};
\begin{scope}[xshift=2cm]
\draw [line,->-=0.5,line width=0.6mm] (8.97,2.2) -- (8.97,2.21);
\draw [line width=0.5mm,smooth,tension=1] plot coordinates {(7,0) (7.5,.75) (9,2) (7.5,3.25) (7,4)};
\node at (7,-.5) {$\cal L$};
\filldraw (8,2) circle (4pt) node[left=0mm] { $\phi(x)$};
\end{scope}
\end{tikzpicture}
\caption{A local operator $\phi(x)$ is said to commute with a topological line $\cal L$ if we can sweep the line past the operator without changing any correlation function. In this case we say that $\phi$ is symmetric under $\cal L$.}
\label{fig:commute}
\end{figure}

There is a closely related, but not identical, action of topological line on operators. 
We can place a local operator near a topological line, and sweep the line past the local operator. 
The hallmark of a non-invertible line is that a local operator can turn into a non-local operator attached to another topological line in this process. 
The most familiar example is that the local, order operator becomes a non-local, disorder operator when it passes through the Kramers-Wannier duality line \cite{Frohlich:2004ef,Chang:2018iay}.

In particular, a local operator $\phi$ is said to commute with a topological line if the correlation function is unchanged as we sweep the line past the local operator as in Figure \ref{fig:commute}.  
This is equivalent to $\widehat{\cal L}|\phi\rangle = \langle{\cal L}\rangle |\phi\rangle$.\footnote{Here $\langle {\cal L}\rangle$ is the quantum dimension of $\cal L$ defined as the eigenvalue of $\widehat{\cal L}$ acting on the identity state, i.e.\ $\widehat{\cal L} |1\rangle = \langle {\cal L}\rangle |1\rangle.$ } 
In this case, we also say the local operator $\phi$ is symmetric under $\cal L$. 
If $\phi$ is a scalar operator, then the RG flow triggered by activating $\phi$ will preserve the non-invertible symmetry generated by $\cal L$.

\begin{figure}
\centering
\begin{tikzpicture}[scale=.5]
\draw (0,0) ++ (180:2 and 1) arc (180:360:2 and 1);
\draw [dashed] (0,0) ++ (0:2 and 1) arc (0:180:2 and 1);
\draw (0,4) ellipse (2 and 1);
\draw (-2,0) -- (-2,4);
\draw (2,0) -- (2,4);
\draw [line,->-=0.75,line width=0.5mm] (0,-1) -- (0,1) -- node[left] {$\cal L$} (0,3);
\node at (0,-1.5) {$|\psi\rangle \in {\cal H}_\cL$};
\node at (4.5,1.5) {$\rightarrow$};
\draw [line,->-=0.5,line width=0.5mm] (8,0) -- node[left] {$\cal L$} (8,4);
\filldraw (8,0) circle (4pt);
\node at (8,-1) {$\psi(x)$};
\node at (11,-1) {};
\end{tikzpicture}
\caption{By quantizing the system on a spatial circle with a topological defect line $\cal L$ inserted at a point in space, we define a defect Hilbert space ${\cal H}_{\cal L}$.  Via the operator-state correspondence, the states in ${\cal H}_{\cal L}$ are mapped to operators living at the end of $\cal L$.}
\label{fig:HL}
\end{figure}
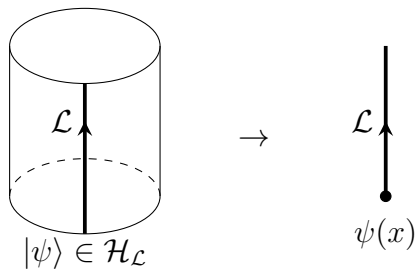

\bigskip\bigskip\centerline{\it Defect Hilbert Space and Spin Selection Rules}\bigskip

When we quantize the 1+1d CFT on a spatial circle $S^1$, we can twist the Hilbert space by a topological defect line that extends in time and intersects with $S^1$ at a point. 
This twisted quantization defines a defect Hilbert space ${\cal H}_{\cal L}$.\footnote{We will use the terms ``defect Hilbert space" and ``twisted Hilbert space" interchangeably. }  
Using the plane-cylinder map, a state $|\psi\rangle $ in ${\cal H}_{\cal L}$ is mapped to a point operator $\psi(x)$ living at the end of the topological line ${\cal L}$.  
This generalizes the ordinary operator-state correspondence between states in the untwisted Hilbert space $\cal H$ and local operators. 
See Figure \ref{fig:HL}.

In a bosonic CFT, every local operator has an integer Lorentz spin, $s\equiv h-\bar h\in \bZ$, due to mutual locality. 
By contrast, point operators $\psi(x)$, which correspond to states in ${\cal H}_{\cal L}$, living at the end of a topological defect line $\cal L$ need not have an integer Lorentz spin.  
In fact, their Lorentz spins are constrained by the fusion category $\cC$ characterizing the topological lines. 
This is called the spin selection rule for the defect Hilbert space ${\cal H}_{\cal L}$. 
In particular, some lines might not even admit an integer spin operator living at its end. 
See   \cite{Chang:2018iay} for derivations and examples of these spin selection rules. 

As two demonstrating examples, consider an invertible $\mathbb{Z}_2$  line $\cal L$ obeying ${\cal L}^2=1$. 
When the $\bZ_2$ symmetry is non-anomalous, the spins in ${\cal H}_{\cal L}$ are constrained to be $s\in \bZ/2$. 
On the other hand, when the $\bZ_2$ is anomalous, the spins in ${\cal H}_{\cal L}$ are constrained to be $s\in \bZ/2+1/4$ \cite{Chang:2018iay,Lin:2019kpn}.

\section{Examples of Non-invertible Symmetries in CFT}
\label{Sec:Examples}

Here we review some of the non-invertible symmetries that we will bootstrap in Section \ref{sec:modular}. 
We also provide several CFT examples that realize these non-invertible symmetries. 
Most of the results in this section are well-known and can be found in, for example, \cite{Rowell:2007dge,Chang:2018iay}. 
We review these results in preparation for comparing them with our numerical bootstrap bounds in Section \ref{sec:modular}.

\subsection{Fibonacci Category}

The unitary Fibonacci category has a single nontrivial line $W$. The fusion rule is
\ie\label{Fib}
W^2 =1+W\,.
\fe
The spin selection rule for the defect Hilbert space ${\cal H}_W$ is \cite{Chang:2018iay}:
\ie\label{Fibspin}
s\in \bZ+ \left\{0 ,\pm \frac 25\right\}
\fe

In rational CFT, there is a finite set of distinguished topological lines that commute with the chiral algebra, known as the Verlinde lines \cite{Verlinde:1988sn,Petkova:2000ip}. 
Both the $(\mathfrak{g}_2)_1$ and $(\mathfrak{f}_4)_1$ WZW models realize the Fibonacci line as its only nontrivial Verlinde line. 
Their central charges are $c=14/5$ and $c=26/5$, respectively. 
Their (bulk) current algebra primary states in the untwisted Hilbert space have the following conformal weights $h,\bar h$:\footnote{Here we label a chiral algebra primary in terms of its conformal dimensions $|h,\bar h\rangle$. More precisely, 
 except for the identity operator, current algebra primaries of fixed conformal weights comprise a multiplet that transforms under a nontrivial representation of the  Lie algebra.}
\ie\label{Fibbulk}
&(\mathfrak{g}_2)_1:~~|0,0\rangle\,,~~~|{2\over5},{2\over5}\rangle \,,\\
&(\mathfrak{f}_4)_1:~~\,|0,0\rangle\,,~~~|{3\over5},{3\over5}\rangle \,.
\fe 
The fusion rule of the current algebra primaries is the same as \eqref{Fib}, with 1 and $W$ replaced by the identity and the nontrivial primary operators, respectively. 
The modular $S$-matrix for both WZW models is
\ie
S = {1\over \sqrt{2+\zeta} }\left(\begin{array}{cc} 1 &~ \zeta \\ \zeta & ~ -1 \\ \end{array}\right)
\fe
 where $\zeta={1+\sqrt{5}\over2}$.  
In diagonal RCFT, the eigenvalue of the action of the $a$-th Verlinde line ${\cal L}_a$ on the $b$-th local chiral algebra primary operator $\phi_b$ is $S_{ab}/S_{1b}$ \cite{Petkova:2000ip}, with $1$ standing for the identity operator. 
In particular, the eigenvalue on the identity local operator (corresponding to choosing $b=1$) is $S_{a1}/S_{11}$, which is the quantum dimension of the line $\langle {\cal L}_a\rangle$. 
In the case of the Fibonacci line in these two WZW models, we have
\ie
\left.\begin{array}{cccc}
(\mathfrak{g}_2)_1 &~~|0,0\rangle~~&~~|{2\over 5}, {2\over 5}\rangle~~& \\
(\mathfrak{f}_4)_1 &|0,0\rangle&|{3\over 5}, {3\over 5}\rangle& \\
\widehat W:&\zeta &-\zeta^{-1}
\end{array}\right.
\fe 
The last line records the eigenvalues of the Fibonacci line operator $\widehat{W}$ on  the corresponding states in the Hilbert space. 

The defect Hilbert space ${\cal H}_W$ twisted by the Fibonacci line $W$ have states with the following conformal weights $h,\bar h$:
\ie
&(\mathfrak{g}_2)_1:~~{\cal H}_W :~~|{2\over 5},0\rangle\,,~~~|0,{2\over 5}\rangle\,,~~~ |{2\over 5} ,{2\over 5}\rangle\,,\\
&(\mathfrak{f}_4)_1:~~\,{\cal H}_W :~~|{4\over 5},0\rangle\,,~~~|0,{4\over 5}\rangle\,,~~~ |{4\over 5} ,{4\over 5}\rangle\,.
\fe
Importantly, the scalar states (i.e.\ those with $h=\bar h$) are not to be confused with the bulk current algebra primaries in \eqref{Fibbulk} in the untwisted Hilbert space ${\cal H}$. 
We see that the spins $s=h-\bar h$ are consistent with the general spin selection rule in \eqref{Fibspin}.

Another simple CFT realizing the Fibonacci category is the tricritical Ising CFT with $c=7/10$. 
There are six Virasoro primary states $|h,\bar h\rangle$:
\ie
|0,0\rangle\,,~~~|{1\over 10}, {1\over 10}\rangle\,,~~~|\frac 35,\frac 35\rangle\,,~~~|\frac 32,\frac 32\rangle\,,~~~|{3\over 80},{3\over 80}\rangle\,,~~~|{7\over 16}, {7\over 16}\rangle \,.
\fe
The actions of the Fibonacci line on these primaries are:
\ie
\left.\begin{array}{ccccccc}
&|0,0\rangle&|{1\over 10}, {1\over 10}\rangle&|\frac 35,\frac 35\rangle&|\frac 32,\frac 32\rangle&|{3\over 80},{3\over 80}\rangle&|{7\over 16}, {7\over 16}\rangle \\
\widehat W:~~&\zeta &-\zeta^{-1}&-\zeta^{-1}&\zeta&-\zeta^{-1}&\zeta
\end{array}\right.
\fe
This non-invertible line was realized as the ``topological symmetry" operator in the anyon chain \cite{Feiguin:2006ydp}, which is a microscopic lattice realization of the tricritical Ising CFT.
 
Consider the RG flow triggered by turning on the local operator $\sigma'$, which is the subleading magnetic deformation, corresponding to $|{7\over 16}, {7\over 16}\rangle$.
Since $\sigma'$ commutes with the Fibonacci line $W$, the latter is preserved along the whole flow. 
On the other hand, $\sigma'$ explicitly breaks the $\bZ_2$ global symmetry. 
The low energy phase of this flow is known to be a gapped phase with two degenerate ground states \cite{Zamolodchikov:1990xc,Mussardo:1992uc,Ellem:1997vz}, which are a direct consequence of the non-invertible line $W$ \cite{Chang:2018iay}.

\subsection{Ising and $\mathfrak{su}(2)_2$ Categories}

The next simplest fusion categories with non-invertible lines are the Ising and $\mathfrak{su}(2)_2$ categories.\footnote{By forgetting the braiding structure, the Ising and the $\mathfrak{su}(2)_2$ fusion categories can be obtained from the unitary modular tensor category (UMTC) for the 2+1d (bosonic) $Spin(\nu)_1$ Chern-Simons theory with $\nu=1,7$ mod 8 and $\nu=3,5$ mod 8, respectively.
 See, for example, \cite{Kitaev:2005hzj} for the TQFT data.}   
Both fusion categories have three simple  lines, $1, \eta, {\cal N}$, obeying the following fusion rule:
\ie
\eta^2=1\,,~~~\eta {\cal N}={\cal N}\eta={\cal N}\,,~~~{\cal N}^2 =1+\eta\,.
\fe
Here $\eta$ is a non-anomalous $\bZ_2$ line and ${\cal N}$ is a non-invertible line.  
As the name suggests, the Ising category is realized in the Ising CFT, in which case $\cal N$ is the Kramers-Wannier duality line \cite{Frohlich:2004ef}. 
While the two categories share the same fusion rule, they differ in their crossing relations, i.e. the $F$-symbols. 
Technically, their Frobenius-Schur indicators have opposite signs.
In the mathematical literature, these two fusion categories are known as the two $\bZ_2$ Tambara-Yamagami categories \cite{TAMBARA1998692}.

The spin selection rules for the defect Hilbert space ${\cal H}_{\cal N}$ of the non-invertible line $\cal N$ are \cite{Chang:2018iay}
\ie\label{Isingspin}
&\text{Ising}:~~s\in {\bZ\over2} \pm {1\over 16}\,,\\
&\mathfrak{su}(2)_2:~~s\in {\bZ\over2} \pm {3\over 16}\,.
\fe
In particular, no state in ${\cal H}_{\cal N}$ has an integer spin.

Let us discuss some examples of 1+1d CFTs realizing these fusion categories. 
The Ising fusion category is realized in the (bosonic)  $\mathfrak{so}(2n+1)_1$ WZW models with $n=0,3$ mod 4.  
The $\mathfrak{su}(2)_2$ fusion category is realized in $\mathfrak{so}(2n+1)_1 $ WZW models with $n=1,2$ mod 4.\footnote{Here $\mathfrak{so}(1)_1=$Ising CFT and $\mathfrak{so}(3)_1=\mathfrak{su}(2)_2$.  For the $\mathfrak{so}(2n)_1$ WZW models, the Verlinde lines are all invertible, which have been considered in \cite{Lin:2019kpn,Lin:2021udi} in the context of the modular bootstrap.} 
In addition, the Ising category is also realized in the Monster CFT \cite{Lin:2019hks}.

Below we focus on the $\mathfrak{so}(2n+1)_1 =(\mathfrak{b}_n)_1$ WZW model, which has three current algebra local primary operators. We denote the corresponding states $|h,\bar h\rangle$ in the Hilbert space $\cal H$ as
\ie\label{Hso}
{\cal H}:~~|0,0\rangle \,,~~~|\frac 12, \frac 12\rangle\,,~~~|{2n+1\over 16} ,{2n+1\over 16}\rangle\,.
\fe
  In the case of $n=0$, i.e. the Ising CFT, these are the identity 1, thermal operator $\varepsilon_{ {1\over2}, {1\over2}}$, and the order operator $\sigma_{{1\over 16},{1\over16}}$, respectively.  
The modular $S$-matrix of the $\mathfrak{so}(2n+1)_1$ WZW model is 
\ie
    \quad
    S = 
    \frac12
    \begin{pmatrix}
        1 & 1 & \sqrt{2}
        \\
        1 & 1 & -\sqrt{2}
        \\
        \sqrt{2} & -\sqrt{2} & 0
    \end{pmatrix}.
\fe

The defect Hilbert space of $\cal N$ consists of the following $\mathfrak{so}(2n+1)_1$ chiral algebra primary states:  
\ie
{\cal H}_{\cal N}:~~|{2n+1\over 16},0\rangle \,,~~~|0,{2n+1\over 16}\rangle\,,~~~|{2n+1\over 16},\frac 12\rangle\,,~~~|\frac 12,{2n+1\over 16}\rangle
\fe
which are consistent with the spin selection rules \eqref{Isingspin}.
Finally, the defect Hilbert space for the (non-anomalous) $\mathbb{Z}_2$ symmetry $\eta$ has the following primary spectrum 
\ie
 \cH_\eta :~ |0, \frac12\rangle, ~~|\frac12, 0\rangle,~~ |\frac{1}{16}, \frac{1}{16}\rangle \,.
 \fe

\section{2+1d TQFT and Anyon Sectors}\label{sec:anyon}

The modular bootstrap equations for a 1+1d CFT with a finite generalized global symmetry can be most compactly described by coupling it to a 2+1d topological quantum field theory (TQFT). 
In this section, we describe this connection for both invertible and non-invertible symmetries. 
The connection will help us understand what a ``charge sector'' means in this context, properly define partition functions in said sectors, and determine their transformation law \eqref{Modular}
which is the central result of this section. 
We emphasize that the final bootstrap results apply to 1+1d CFT on its own, without coupling to any higher dimensional TQFT; the latter is introduced as a trick for formulating the bootstrap equations.

We start with a general discussion on symmetry twists and different charge sectors of a 1+1d CFT, and later relate these to sectors of a 2+1d system. 
In the presence of a finite invertible global symmetry, the partition function can be refined by turning on flat background gauge fields, and the mapping class group of the gauge bundle relates partition functions with different background gauge configurations.  
The analysis of such a system of twisted partition functions refines the modular bootstrap into constraints on various charged and twisted sectors.
For a finite group symmetry $G$ with an 't Hooft anomaly $\omega\in H^3(G,U(1))$ in a 1+1d CFT, each defect Hilbert space ${\cal H}_{[g]}$ is labeled by a conjugacy class $[g]$. 
We can further grade the states in ${\cal H}_{[g]}$ by the action of the centralizer $\Cen_G(g)$ into different charge sectors. 
More specifically, each defect Hilbert space ${\cal H}_{[g]}$ is further decomposed in subsectors ${\cal H}_{[g],\rho}$, each labeled by $([g],\rho)$, where  $\rho$ is an irreducible projective representations of  $\Cen_G(g)$, with projectivity specified by $\omega$ and $[g]$.

Physically, $g$ corresponds to the insertion of a time-like $g$-defect to twist the periodic boundary condition on the spatial $\S^1$, and $\Cen_G(g)$ comprises the consistent space-like operator insertions.  
The modular bootstrap refined by a finite symmetry (possibly with anomalies) has been considered in \cite{Lin:2019kpn,Pal:2020wwd,Lin:2021udi,Grigoletto:2021zyv,Cao:2021euf,Lanzetta:2022lze}.

There is also a 2+1d interpretation of these different sectors ${\cal H}_{[g],\rho}$ in a 1+1d CFT.
We  couple the 1+1d CFT with a $G$ symmetry and an anomaly $\omega$ to a 2+1d $G$ gauge theory with a twist $\omega\in H^3(G,U(1))$. 
This finite group gauge theory is a TQFT, which is also known as the Dijkgraaf-Witten theory \cite{Dijkgraaf:1989pz}, or the twisted quantum double $D^\omega (G)$ (see also \cite{Dijkgraaf:1990ne,deWildPropitius:1995cf,Hu:2012wx}). 
After the coupling, the CFT becomes a boundary condition of the 2+1d TQFT. 
Each simple anyon line in this 2+1d TQFT is labeled by $([g], \rho)$. 
Each sector ${\cal H}_{[g],\rho}$ of the 1+1d CFT now becomes the Hilbert space of point operators where the anyon can terminate on the 1+1d boundary.

For a non-invertible symmetry, there is a similar notion of sectors---we can insert time-like defects to twist the periodic boundary condition, and then decompose the defect Hilbert spaces into irreducible representations of the \textit{tube algebra} \cite{ocneanu1994chirality,evans1995ocneanu} implemented by the \textit{lasso} actions of these non-invertible lines. 
We refer the reader to \cite{Chang:2018iay} for the definition of the lasso action and to \cite{Williamson:2017uzx,Aasen:2020jwb,Lin:2022dhv}  for a detailed physics discussion of the tube algebra.\footnote{Note that each sector can involve multiple defect Hilbert spaces because lassos generally do not preserve the twisting. This is analogous to the situation for invertible symmetries where two defect Hilbert spaces ${\cal H}_g$ and ${\cal H}_{g'}$ can be mapped to each other by a symmetry action if $g,g'$ are in the same conjugacy class.}
It is a remarkable fact due to \cite{evans1995ocneanu,Izumi:2000qa,Muger}\footnote{See also \cite{Williamson:2017uzx,Aasen:2020jwb} in the contexts of tensor networks and statistical lattice models and \cite{Lin:2022dhv} for a continuum field theory explanation.} that the irreducible representations of the tube algebra are in one-to-one correspondence with the simple objects of the Drinfeld center (quantum double), whose modular data conveniently gives the modular relations among the different sectors.

\begin{figure}[h]
    \centering
    \raisebox{-73pt}{\begin{tikzpicture}[scale=1.5]
        \draw [<->] (1,4.2) -- (3,4.2);
        \draw (2,4.2) node[above]{$I$};
        \draw (1.35,0.8) node[below]{\small \color{red!75!DarkGreen} $\TV_\cC$};
        \draw[color=DarkGreen] (0,0) -- (0,3) -- (1,4) -- (1,1) -- cycle;
        \draw (0,0) node[below]{\color{DarkGreen} $\B_Q$};
        \draw [color=red!75!DarkGreen, thick,
        decoration = {markings, mark=at position 0.6 with {\arrow[scale=1]{stealth}}}, 
        postaction=decorate] (0.5,2) -- (1.5,2) node[below]{$\mu$} -- (2.5,2);
        \draw [fill=DarkGreen] (0.5,2) circle (0.04) node [below] {\color{DarkGreen} $\widetilde{\cO}$};
        \draw [color=blue!70!green, thick, 
        decoration = {markings, mark=at position 0.5 with {\arrow[scale=1]{stealth}}}, 
        postaction=decorate] (2.5,2) -- (2.5,3) node[right]{$a$} -- (2.5,3.75);
        \draw [fill=blue!70!green] (2.5,2) circle (0.04) node [below] {\color{blue!70!green} $x$};
        \draw[color=blue!70!green, preaction={draw=white,line width=3pt}] (2,0) -- (2,3);
        \draw[color=blue!70!green] (2,3) -- (3,4) -- (3,1) -- (2,0);
        \draw (2,0) node[below]{\color{blue!70!green} $\D$};
    \end{tikzpicture}}
    \quad $=$ ~
    \raisebox{-73pt}{\begin{tikzpicture}[scale=1.5]
        \draw [color=blue!70!green, thick, 
        decoration = {markings, mark=at position 0.5 with {\arrow[scale=1]{stealth}}}, 
        postaction=decorate] (2.5,2) -- (2.5,3) node[right]{$a$} -- (2.5,3.75);
        \draw [fill=black] (2.5,2) circle (0.04) node [below] {$\cO$};
        \draw(2,0) -- (2,3) -- (3,4) -- (3,1) -- cycle;
        \draw (2,0) node[below]{$Q$};
        \end{tikzpicture}}
    \caption{A defect operator $\cO\in\cH_a$ of the 2d theory $Q$ is composed of a bulk anyon $\mu\in\ZC$ and a pair of boundary (defect) operators.  (Figure reproduced from \cite{Lin:2022dhv}.)
    }
\label{Fig:DefectOperators}
\end{figure}
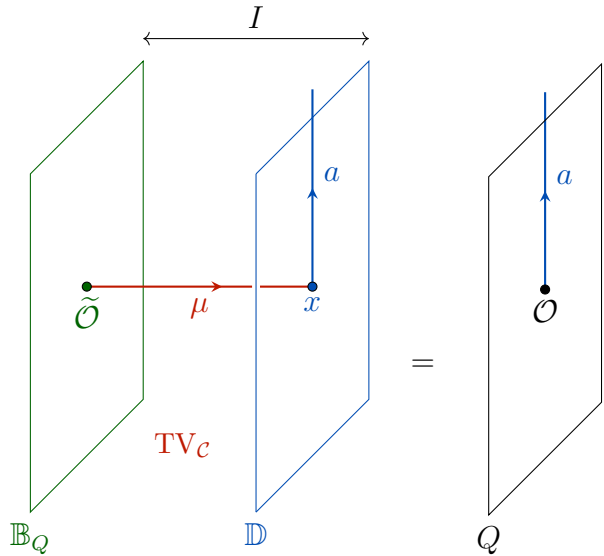

This correspondence, as explained in \cite{Lin:2022dhv},  becomes quite intuitive in the slab construction \cite{Gaiotto:2020iye,Freed:2022qnc} depicted in Figure~\ref{Fig:DefectOperators}.  
We start with a 1+1d CFT $Q$ with a (non-invertible) finite symmetry described by a fusion category $\cal C$, whose simple objects are the simple topological lines labeled by $a$. 
Given any fusion category $\cal C$, there is a corresponding 2+1d TQFT known as the Turaev-Viro TQFT $\TV_\cC$. 
The unitary modular tensor category  (UMTC) data of $\TV_\cC$ are given by the Drinfeld center $\mathcal{Z}({\cal C})$, whose simple objects, labeled by $\mu$, are the anyon lines. 
The Turaev-Viro TQFT can be viewed as a non-invertible generalization of a finite group gauge theory based on $\cC$, i.e.\ it is a 
2+1d $\cal C$ gauge theory (see e.g.\ \cite{Kaidi:2021gbs} for discussions on this interpretation). 
In particular, the CFT $Q$ can be coupled to a 2+1d Turaev-Viro TQFT $\TV_\cC$ by gauging $\cal C$ (see e.g. \cite{Thorngren:2019iar,Lin:2022dhv}). 
This is a non-invertible generalization of the setup where a $G$ gauge theory in one-higher dimension is coupled to a QFT with $G$ global symmetry, in which case $a$ is given by a conjugacy class $[g]$ in $G$  and $\mu \in ([g],\rho)$.  
Recently, the Turaev-Viro TQFT has also been referred to as the Symmetry TFT for non-invertible symmetries \cite{Kaidi:2022cpf,Kaidi:2023maf}.

Now, the 1+1d CFT $Q$ becomes a boundary condition $\bB_Q$ for the TQFT $\TV_\cC$.  To recover the original CFT, one places the TQFT   on a slab with a Dirichlet boundary condition $\D$ on one side, and $\B_Q$ on the other \cite{Gaiotto:2020iye,Freed:2022qnc}.  
The topological Dirichlet boundary condition $\D$ has the distinguished property that the set of topological lines on it form the original fusion category $\cC$. 
We can define a (forgetful) map $F:{\cal Z}(\cC)\to \cC$  by the physical process of bringing   a bulk anyon $\mu$ to the boundary in parallel fashion to produce a boundary topological line  $F(\mu)$, which is generally semi-simple (a sum of simple defects in $\cC$).
Each point-like operator in $Q$ becomes a point-like operator on the non-topological boundary $\B_Q$ attached to a bulk anyon line.  Because all symmetry operations happen on the Dirichlet boundary $\D$, the anyons label the irreducible representations of the symmetry algebra $\text{Tube}(\cC)$.

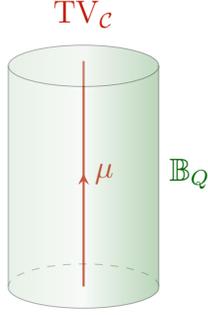
\begin{figure}[t]
\centering
\raisebox{-60pt}{\begin{tikzpicture}
\draw[color=red!75!DarkGreen, thick, decoration = {markings, mark=at position 0.5 with {\arrow[scale=1]{stealth}}}, postaction=decorate] (0,0) -- (0,1.5)node[right]{$\mu$} -- (0,3);
\draw (0,3.3) node[above]{\color{red!75!DarkGreen} $\TV_\cC$};
\draw (1,1.5) node[right]{\color{DarkGreen} $\B_Q$};
\node[style={draw, shape=cylinder, opacity=0.3, aspect=2, minimum height=+3.5cm,
minimum width=+2cm, left color=DarkGreen!30, right color=DarkGreen!90, middle color=DarkGreen!10,
shape border rotate=90, append after command={
  let \p{cyl@center} = ($(\tikzlastnode.before top)!0.5! (\tikzlastnode.after top)$),
      \p{cyl@x}      = ($(\tikzlastnode.before top)-(\p{cyl@center})$),
      \p{cyl@y}      = ($(\tikzlastnode.top)       -(\p{cyl@center})$)
  in (\p{cyl@center}) 
  edge[draw=none, opacity=0.2, fill=DarkGreen!10, to path={
    ellipse [x radius=veclen(\p{cyl@x})-1\pgflinewidth,
             y radius=veclen(\p{cyl@y})-1\pgflinewidth,
             rotate=atanXY(\p{cyl@x})]}] () }}] at (0,1.25) {};
\draw[color=DarkGreen, opacity=0.3, dashed] (1,0) arc [start angle=0, end angle=180, x radius=1, y radius=0.3];
\end{tikzpicture}}
    \caption{The 2+1d TQFT $\TV_\cC$ quantized on a spatial disk with boundary condition $\B_Q$ and the anyon line $\mu$ inserted at the origin. We denote the Hilbert space of this 2+1d system on a disk by ${\cal V}_\mu$. 
    We denote the partition function over this Hilbert space by $Z^\text{3d}_\mu(\tau,\bar \tau)$.  (Figure reproduced from \cite{Lin:2022dhv}.)}
    \label{Fig:SolidTorus}
\end{figure}

For each anyon $\mu$, we consider the Hilbert space ${\cal V}_\mu$ from quantizing the TQFT $\TV_\cC$ on a spatial disk with boundary condition $\B_Q$ and the anyon line defect inserted at the origin, as shown in Figure~\ref{Fig:SolidTorus}. 
The defect Hilbert spaces ${\cal H}_a$ of the 1+1d CFT $Q$ are related to the anyon sectors ${\cal V}_\mu$ of the 2+1d TQFT as \cite{Lin:2022dhv}
\ie\label{HV}
{\cal H}_a = \bigoplus_\mu W^\mu_a \otimes {\cal V}_\mu
\fe
where $W^\mu_a = \text{Hom}_\cC (F(\mu) , a)$ is the vector space of $\D$-boundary operators $x$ at the junction of $\mu$ and $a$.

Of special importance to us is the anyon sector ${\cal V}_\mathbf{1}$ corresponding to the trivial anyon $\mu = \mathbf{1}$.\footnote{We use boldface $\mathbf{1}$ for the trivial anyon in the 2+1d TQFT, to distinguish from the trivial defect line in the 1+1d CFT.}
We claim that all the $\cC$-symmetric states in the untwisted Hilbert space ${\cal H} = {\cal H}_1$ of the 1+1d CFT arise from ${\cal V}_\mathbf{1}$ via \eqref{HV}. 
These are states $|{\cal O}\rangle\in {\cal H}$ such that $\widehat{\cal L}_b |{\cal O}\rangle = \langle{\cal L}_b\rangle |{\cal O}\rangle $ for all $b\in \cC$, which  correspond to local operators $\cal O$ that commute with the entire $\cC$ (see Figure \ref{fig:commute}).
To see this claim, we encircle a local operator $\cal O$ by a topological line ${\cal L}_b$ of $\cC$  on the right side of Figure \ref{Fig:DefectOperators} (with $a=1$). 
Assuming that $\cal O$ belongs to the $\cV_\mathbf{1}$ sector, we can replace the original 1+1d configuration with the 2+1d configuration on the left of the figure with $\mu= \mathbf{1}$. 
The junction operator $x  \in W^{\mu=\mathbf{1} }_{a=1}$  is trivial, therefore the topological line ${\cal L}_b$ on the $\mathbb{D}$-boundary acts on $x$ by its quantum dimension $\langle{\cal L}_b\rangle$. 
It follows that  ${\cal O}$ commutes with all the lines of $\cC$  in the 1+1d CFT.  
Conversely, all the lines in $\cC$ act by their quantum dimensions only if $\mu=\mathbf{1}$.\footnote{Otherwise, the half-linking matrix is degenerate, contrary to what was proven in \cite[Appendix~C]{Lin:2022dhv}. Note that the projection operator 
to the $\cC$-symmetric sector of $\cH$ is given in \cite[(4.26)]{Lin:2022dhv} such that $P\cH = \cV_\mathbf{1}$.}
We conclude that all the $\cC$-preserving deformations arise from the scalar states in ${\cal V}_\mathbf{1}$, which, if relevant $\Delta<2$, would potentially destabilize the gapless phase even when the $\cC$ symmetry is imposed.

Let $Z^\text{3d}_\mu(\tau, \bar\tau)$ be the partition function for ${\cal V}_\mu$. Explicitly, it is the TQFT partition function on a solid torus of modulus $\tau$ with boundary condition $\B_Q$ and a $\mu$ anyon looping the non-contractible cycle. 
 Under modular transformations, $Z^\text{3d}_\mu$ transforms as
\ie\label{Modular}
    Z^\text{3d}_\mu(\tau+1, \bar\tau+1) &= T_{\mu\mu} Z^\text{3d}_\mu(\tau, \bar\tau),
    \quad
    Z^\text{3d}_\mu(-1/\tau, -1/\bar\tau) &= \sum_{\nu\in {\cal Z}({\cal C})} S_{\mu\nu} Z^\text{3d}_\nu(\tau, \bar\tau),
\fe 
where $T$ and $S$ are the modular matrices  of  $\ZC$ and the sum is over the simple objects (i.e.\ the anyons) of ${\cal Z}({\cal C})$.\footnote{There are different conventions for $S$ and $T$ up to overall factors. Here we fix our convention by choosing $S^2 = (ST)^3 = 1$.
}
In Section \ref{sec:modular}, we will use the modular properties of these partition functions to bootstrap 1+1d CFT with a non-invertible global symmetry.  
We comment that \eqref{Modular} can in principle be derived from a purely 1+1d reasoning, such as in \cite{Lin:2019kpn,Pal:2020wwd,Lin:2021udi} for invertible symmetries, but the computation can in practice be quite intractable. Moreover, the 2+1d construction provides a much more conceptual derivation. 
These partition functions have recently been studied in \cite{Ji:2019eqo,Ji:2019jhk,Ji:2021esj,Chatterjee:2022tyg,Chatterjee:2022jll} in the context of topological order and categorical symmetries.

\bigskip\bigskip\centerline{\it Example: Ising CFT}\bigskip

We demonstrate the above general discussion using the simplest example of Ising CFT. 
The modular data of the Ising CFT are given by
 \ie\label{IsingMD}
     S^\text{Ising} = 
     \frac12
     \begin{pmatrix}
         1 & 1 & \sqrt{2}
         \\
         1 & 1 & -\sqrt{2}
         \\
         \sqrt{2} & -\sqrt{2} & 0
     \end{pmatrix},
     \quad
     T^\text{Ising} =
     e^{-\frac{2\pi i}{48}}
     \begin{pmatrix}
         1
         \\
         & -1
         \\
         & & e^{\frac{\pi i}{8}}
     \end{pmatrix},
 \fe 
 where we order the primaries by $1_{0,0}, \varepsilon_{\frac 12,\frac12}, \sigma_{{1\over 16},{1\over 16}}$ with conformal weights shown in their subscripts. 
We denote the corresponding $c=\frac 12$ torus characters as $\chi_h^\text{Ising}$  with $h=0,\frac 12, {1\over 16}$.

Before we discuss the non-invertible duality defect, let us first focus on the non-anomalous $\bZ_2$ symmetry, i.e.\ $\cC = \text{Vec}(\bZ_2)$, as a warm-up. 
We denote the two simple objects of $\text{Vec}(\bZ_2)$ as $1,\eta$. 
 The Drinfeld center $\ZC$ describes the data for the 2+1d $\mathbb{Z}_2$ gauge theory, which is the low-energy phase of the 2+1d toric code. For this reason, we will refer to it as the toric code UMTC. The modular matrices for the toric code are (see, for example, \cite{Kitaev:2005hzj})
\ie\label{TCMD}
    S^\text{TC} = \frac12
    \begin{pmatrix}
        1 & 1 & 1 & 1
        \\
        1 & 1 & -1 & -1
        \\
        1 & -1 & 1 & -1
        \\
        1 & -1 & -1 & 1
    \end{pmatrix},
    \quad
    T^\text{TC} = \text{Diag}(1, 1, 1, -1).
\fe 
We label the four anyons as $\mathbf{1},e, m,f$, whose spins $h$ are $0,0,0,\frac 12 \pmod 1$. The forgetful map $F:{\cal Z}(\cC)\to \cC$ acts as $F(\mathbf{1})=F(e)=1$ and $F(m)=F(f)=\eta$.

The trivial anyon sector $\mu = \mathbf{1}$ corresponds to the $\bZ_2$-even states in the untwisted Hilbert space ${\cal H}^+$, i.e.\ the identity $1_{0,0}$ and $\varepsilon_{\frac 12,\frac 12}$, as well as their descendants,
\ie 
    Z^\text{3d}_\mathbf{1} = \chi_0^\text{Ising} \bar\chi_0^\text{Ising} + \chi_{\frac12}^\text{Ising} \bar\chi_{\frac12}^\text{Ising}\,.
\fe  
Next, $\mu = e$ and $\mu = m$ are on equal footing, corresponding to $\bZ_2$-odd states in the untwisted Hilbert space ${\cal H}^-$, i.e.\ the spin/order operator and its descendants, and $\bZ_2$-even states in the twisted Hilbert space ${\cal H}_\eta^+$, i.e.\ the disorder operator and its descendants,
\ie 
    Z^\text{3d}_e = Z^\text{3d}_m = \chi^\text{Ising}_{\frac{1}{16}} \bar\chi^\text{Ising}_{\frac{1}{16}}.
\fe 
(The $\bZ_2$-charge in the twisted Hilbert space ${\cal H}_\eta$ can be read off from the Lorentz spin $h-\bar h$ by a spin selection rule \cite{Chang:2018iay,Lin:2019kpn}.)
Finally, $\mu = f$ corresponds to $\bZ_2$-odd states in the twisted Hilbert space ${\cal H}_\eta^-$, i.e.\ the chiral free fermions and their descendants,
\ie \label{Zf}
    Z^\text{3d}_f = \chi^\text{Ising}_0 \bar\chi^\text{Ising}_{\frac12} + \chi^\text{Ising}_{\frac12} \bar\chi^\text{Ising}_0.
\fe 
Using  \eqref{IsingMD}, one can immediately verify that $Z^\text{3d}_\mu$ transform according to the toric code modular data \eqref{TCMD}.
To summarize, we have the following identification between the defect Hilbert spaces in 1+1d and the anyon sectors in 2+1d [\cite{Lin:2019kpn}, Sec. 2.4]:\footnote{To compare the notations ${\cal H}^\pm, {\cal H}_\eta^\pm$ here with ${\cal H}_{[g], \rho}$ earlier this section for a general finite group $G$, we have $[g]=1,\eta$ and $\rho =\pm$.}
\ie
&{\cal H}^+ \cong {\cal V}_\mathbf{1}:~|0,0\rangle\,,\\
&{\cal H}^- \cong {\cal V}_e:~|{1\over 16},{1\over 16}\rangle\,,\\
&{\cal H}^+_\eta \cong {\cal V}_m:~|{1\over 16},{1\over 16}\rangle\,,\\
&{\cal H}^-_\eta \cong {\cal V}_f:~|\frac 12,0\rangle, |0,\frac 12\rangle\,.
\fe

Now, let us incorporate the non-invertible duality defect $\cN$ and choose $\cC$ to be the  Ising fusion category. 
The defect Hilbert spaces for $1,\eta, \cN$ in the 1+1d CFT are \cite{Hauru:2015abi,Chang:2018iay,Lin:2019hks}
\ie
&{\cal H}:~~~|0,0\rangle\,,~~~|\frac 12, \frac 12\rangle\,,~~~|{1\over16},{1\over 16}\rangle\,,\\
&{\cal H}_\eta:~~ |{1\over 16}, {1\over 16}\rangle\,,~~~|\frac 12,0\rangle \,,~~~|0,\frac12\rangle\,,\\
&{\cal H}_\cN:~|0,{1\over 16}\rangle\,,~~~|{1\over 16}, 0\rangle\,,~~~|\frac 12,{1\over 16}\rangle \,,~~~|{1\over 16}, \frac 12\rangle\,.
\fe
Next, we couple the 1+1d Ising CFT to the 2+1d TQFT described by the Drinfeld center of the Ising fusion category. 
The latter  is  $\text{Ising} \boxtimes \overline{\text{Ising}}$, i.e.\ the 2+1d double Ising TQFT.  
The 9 anyons can be labeled as $\mu = (a, \bar b)$, with $a,b=1,\eta,\cN$.   
They lead to 9 different anyon sectors ${\cal V}_{(a,\bar b)} $, each containing exactly one primary state $|h_a,\bar h_{\bar b}\rangle$  with $h_1=0, h_\eta=\frac12, h_\cN= {1\over16}$ \cite{Ji:2019jhk}. 
The corresponding partition functions are  
\ie
Z^\text{3d}_{(a, \bar b)} = \chi^\text{Ising}_{h_a} \bar\chi^\text{Ising}_{\bar h_{\bar b}}\,,
\fe
which manifestly transform under the modular data of $\ZC$, since $S^{\cZ(\text{Ising})} = S^\text{Ising} \otimes S^\text{Ising}$ and $T^{\cZ(\text{Ising})} = T^\text{Ising} \otimes \bar T^\text{Ising}$.  
In particular,  ${\cal V}_\mathbf{1} $ (where $\mathbf{1}=(1,1))$ only contains the identity module, so there is no nontrivial scalar primary in the Ising CFT that is invariant under the full Ising category.

Note that there are in total 10 primary states in the untwisted and twisted Hilbert spaces in 1+1d, but there are only 9 primary states in total in all the anyon sectors in 2+1d. 
More specifically, in 1+1d, the spin/order operator $\sigma_{{1\over 16} , {1\over 16}}$ is  a local operator in the  untwisted Hilbert space ${\cal H}$, while the disorder operator $\mu_{{1\over16}, {1\over 16}}$ is a non-local operator in the  twisted Hilbert space ${\cal H}_\eta$. 
They are mapped to each other by the lasso action in \cite{Chang:2018iay}:
\ie
    \label{Lasso}
	\raisebox{-2.5em}{
    \begin{tikzpicture}[scale=.5]
    \draw [thick, dashed] 
    (0,2) -- (0,3) node[right] {$\eta$} -- (0,4);
    \draw [thick] (0,0) circle (2);
     \draw [fill=black] (0,0) circle (.1);
   \draw   (0,-.8) node   {$\sigma_{{1\over 16},{1\over 16}}$};
    \draw (-2,0) node[left] {$\cN$};
        \draw (4,0) node[left=3pt] {$=$};
            \draw [thick, dashed]     (5,0) -- (5,4)  ;
                 \draw [fill=black] (5,0) circle (.1);
                    \draw   (5,-.8) node   {$\mu_{{1\over 16},{1\over 16}}$};
                        \draw (5.5,3) node {$\eta$};
    \end{tikzpicture}}
\fe
After we couple the 1+1d Ising CFT to the 2+1d double Ising TQFT, they are identified as a single state in the anyon sector $ {\cal V}_{(\cN, \bar\cN)}$ in 2+1d.

\bigskip\bigskip\centerline{\it General RCFTs}\bigskip

For a general diagonal rational CFT associated with a UMTC $\cC$, we can consider the set of topological lines that commute with the extended chiral algebra. 
The fusion category of these lines, known as the Verlinde lines \cite{Verlinde:1988sn,Petkova:2000ip,Chang:2018iay}, is obtained from $\cC$ by forgetting the braiding structure.

We can gauge $\cC$ by coupling the (non-chiral) RCFT  to the 2+1d Turaev-Viro TQFT, which in this case is given by $\ZC \cong \cC \boxtimes \bar\cC$ \cite{Muger},
where $\bar\cC$ denotes the orientation reversal of the UMTC $\cC$. 
It is important to note that in writing this, $\cC$ must be understood as a UMTC and not just a fusion category, so that $\cC$ and $\bar\cC$ are generically different, even if they are identical as fusion categories.

We can unfold the above configuration to have the 2+1d TQFT  $\cC$ in the bulk on an interval, sandwiched by the chiral and anti-chiral RCFTs as the boundary conditions at the two ends. 
Each primary local operator $\phi_a$ of the original non-chiral RCFT is now represented by an anyon line $a$ of $\cC$ stretched between the two boundaries. 
When we fold again, we find that each anyon sector $(a,\bar b)$ of $\mathcal{Z}(\cC)  =\cC \boxtimes \bar \cC$  contains exactly one primary state $|h_a, \bar h_{\bar b}\rangle$, i.e. 
\ie\label{ZRCFT}
   Z^\text{3d}_{(a,\bar b)} = \chi_{h_a} \bar\chi_{\bar h_{\bar b}}.
\fe 
From the unfolded frame, it is also clear that the forgetful map in this case is given by $F( a,\bar b) = a\otimes \bar b$.  
Using \eqref{HV}, we have 
\ie\label{Hc}
{\cal H}_c = \sum_{a,b\in \cC} N^a_{bc}\, {\cal V}_{(a,\bar b)}
\fe
where $N^a_{bc}$ is the fusion coefficient of the RCFT. 
Note that for fixed $a, b$, if $c \neq c'$ are such that $N^a_{bc}$ and $N^a_{bc'}$ are both nonzero, then this means that the corresponding states in $\cH_c$ and $\cH_{c'}$ are mapped to each other by a lasso action.

We are interested in scalar states of the RCFT that preserve the full non-invertible symmetry $\cC$. 
When coupled to the TQFT ${\cal Z}(\cC)$, they reside in the trivial anyon sector $(a,\bar b) = (1,1)$ (i.e.\ ${\cal V}_\mathbf{1}$). 
In the RCFT case, ${\cal V}_\mathbf{1}$ has a scalar gap determined by the first nontrivial Virasoro primary in the chiral algebra.  
In the case of a  Kac-Moody algebra, this gap is 2, realized by the $\sum_A j^A \bar j^A$ current bilinears, which is a $\cC$-preserving marginal scalar Virasoro primary. (Here $A$ denotes the adjoint index of the associated Lie algebra.)
For a $W(2,s)$ algebra, this gap is $2s$.  In particular, any rational CFT with no spin-one current is stable against any $\cC$-preserving deformation, i.e.\ it is a $\cC$-protected gapless phase.\footnote{
When spin-one currents are present, one can consider the coset which has a different modular tensor category.
}

It is worth emphasizing that the Verlinde lines $\cC$ are not the only symmetries an RCFT enjoys. 
For instance, the continuous Lie group symmetry of the WZW model doesn't commute with the current algebra and is not part of the Verlinde lines. 
In that case, the symmetry-preserving scalar gap generally does not follow the above simple rule.

\section{Constraints from the Modular Bootstrap}\label{sec:modular}

\subsection{Modular Bootstrap and the Linear Functional Method}

Our goal is to derive rigorous bounds on the scaling dimensions of states in various twisted and untwisted Hilbert spaces ${\cal H}_a$ of a general 1+1d CFT with a fusion category symmetry $\cC$.  
To achieve this, we couple the 1+1d CFT to a 2+1d TQFT as in Section \ref{sec:anyon}. 
The partition functions $Z^\text{3d}_\mu$ for the resulting 2+1d system obey a simple modular property as in \eqref{Modular}, which we can use as the bootstrap equations to derive rigorous bounds on the anyon sectors ${\cal V}_\mu$ in 2+1d. 
The relation \eqref{HV} in turn translates these bounds on ${\cal V}_\mu$ into those on ${\cal H}_a$ for the original 1+1d CFT.

To implement the numerical bootstrap, 
we decompose  each partition function $Z^\text{3d}_\mu$ into Virasoro characters,
\ie\label{CharacterExpansion}
    Z^\text{3d}_\mu(\tau, \bar\tau) = \sum_{ (h, \bar h)\in {\cal H}_\mu} n_{\mu; h, \bar h} \chi_h(\tau) \chi_{\bar h}(\bar\tau),
\fe
where the coefficients $n_{\mu; h, \bar h}\in \mathbb{Z}_{\ge 0}$ are non-negative integers by unitarity.
In our numerical bootstrap, we assume $c > 1$, so that the Virasoro characters are non-degenerate
\ie 
    \chi_h(\tau) = \frac{e^{2\pi i \tau (h-c/24)}}{\eta(\tau)},  ~~~(h\neq 0)
\fe 
except for the vacuum character $h=0$
\ie 
    \chi_0(\tau) = \frac{q^{h-c/24}}{\eta(\tau)} (1-q),
\fe
where $q = \exp(2\pi i \tau)$.

Using the character expansion, we can write the modular covariance equation \eqref{Modular} in the form
\ie\label{ModularX}
    0 &= \sum_{\nu\in{\cal Z}({\cal C})} \sum_{(h, \bar h) \in \cH_\nu} n_{\nu; h, \bar h} X_{\mu,\nu; h,\bar h}(\tau, \bar\tau), \quad \forall \mu\in\text{anyons},
\fe 
where 
\ie
    X_{\mu,\nu; h,\bar h}(\tau, \bar\tau) &= I_{\mu\nu} \chi_h(-1/\tau) \chi_{\bar h}(-1/\bar\tau) - S_{\mu\nu} \chi_h(\tau) \chi_{\bar h}(\bar\tau),
\fe 
where $I_{\mu\nu}$ is the identity matrix. 

We are now ready to run the modern numerical bootstrap \cite{Rattazzi:2008pe,Poland:2018epd}, sometimes also called the linear functional method, to produce nontrivial bounds. 
 In a nutshell, the idea is to postulate that the conformal weights $(h, \bar h)$ contributing to $Z^\text{3d}_\mu$ have putative support $\cP_\mu$, and derive contradictions with \eqref{Modular}, \eqref{CharacterExpansion}, 
to rule out the collection $\{\cP_\mu\}$.  First, the $T$-transform in \eqref{Modular} determines the fractional spin $h - \bar h \mod \bZ$ of each anyon sector.  Then, we act with a linear functional, typically chosen to be
\ie 
    \alpha_\mu = \sum_{m, n = 0}^\Lambda \alpha_\mu^{m,n} \partial_\tau^m \partial_{\bar\tau}^n|_{\tau = -\bar\tau = i}, \quad \alpha_\mu^{m,n} \in \bR,
\fe
on \eqref{ModularX} to get
\ie 
    0 = \sum_{\mu,\nu\in{\cal Z}({\cal C})} \sum_{(h, \bar h) \in \cH_\nu} n_{\nu; h, \bar h} \, \alpha_\mu[ X_{\mu,\nu; h,\bar h} ],
\fe 
and search in the space of $\alpha_\mu^{m,n}$ to make $\alpha_\mu[ X_{\mu,\nu; h,\bar h} ]$ non-negative on the entire $\cP_\mu$, 
and hence arrive at a contradiction.  
More details can be found in \cite{Lin:2019kpn}.   If such a functional exists, then the putative spectrum is ruled out.  Iterating this procedure produces various constraints, such as bounds on the gap in the spectrum of symmetry-preserving ($\mu = \mathbf{1}$) scalar primaries.  The numerical bounds in this paper are all obtained at derivative order $\Lambda = 19$ and with spin truncation $s^\text{max} = 38$.  The search for a linear function utilizes the semi-definite programming solver SDPB \cite{Simmons-Duffin:2015qma,Landry:2019qug}.\footnote{We use the following SDPB parameter settings: $\texttt{precision=768}$, \, $\texttt{initialMatrixScalePrimal=1e-10}$, \, $\texttt{initialMatrixScaleDual=1e-10}$, \, $\texttt{maxComplementarity=1e-30}$, \, $\texttt{feasibleCenteringParameter=0.1}$, \, $\texttt{infeasibleCenteringParameter=0.3}$, \, $\texttt{stepLengthReduction=0.7}$.
}

\subsection{Bounds on Symmetry-Protected Gapless Phases}\label{sec:scalar}

Given a fusion category $\cC$,  we are particularly interested in the space of CFTs that are stable under any $\cC$-preserving local deformation. 
This can be thought of as a robust $\cC$-protected gapless phase. 
Such CFTs have no relevant scalar primary operators in the symmetry-preserving $\mu = \mathbf{1}$ anyon sector.  
The space of stable CFTs with a $\mathbb{Z}_N$ global symmetry was studied in \cite{Lin:2019kpn,Lin:2021udi} by the modular bootstrap.

In the modular bootstrap,  we choose $\{\cP_\mu\}$ to be the unitarity bound $h, \bar h \ge 0$ for all $\cV_{\mu \neq \mathbf{1}}$,  as well as for the non-scalar subsector $h\neq\bar h$  of $\cV_\mathbf{1}$. 
For the scalar subsector of $\cV_\mathbf{1}$, we further impose a gap $h = \bar h \ge \Delta/2$.  For any given central charge $c $, we numerically search a critical $\Delta_*$ such that $\Delta > \Delta_*$ is ruled out while $\Delta < \Delta_*$ is not.  This $\Delta_*$ is then a rigorous bound on the symmetry-preserving scalar gap $\Delta$.  If $\Delta_* < 2$, then a symmetry-protected gapless phase does not exist.

\begin{figure}[h!]
    \centering
    \subfloat{
    \includegraphics[height=.37\textwidth]{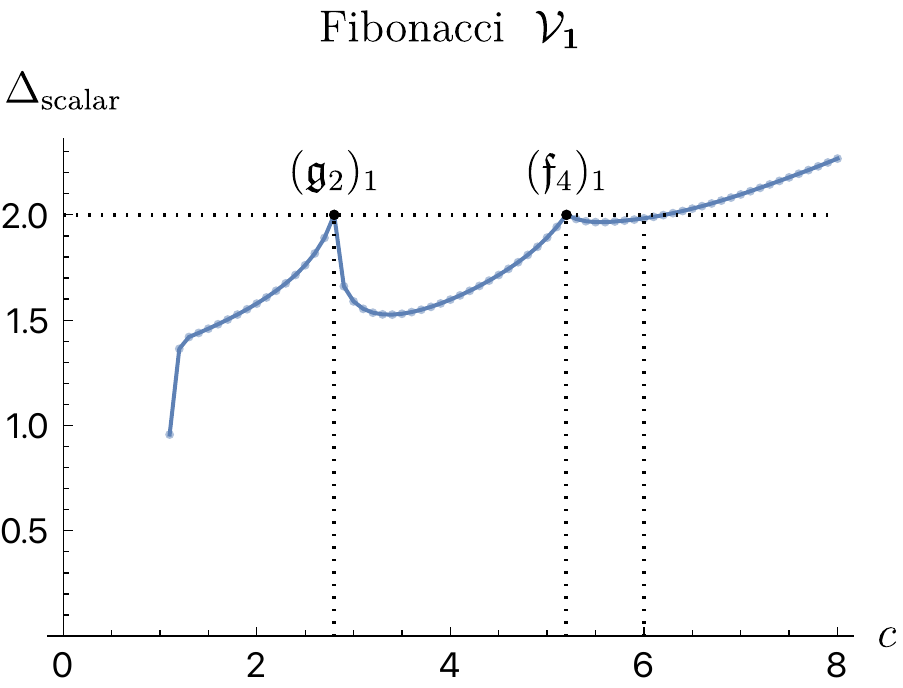}
    }
    \\
    ~
    \\
    \subfloat{
    \includegraphics[height=.37\textwidth]{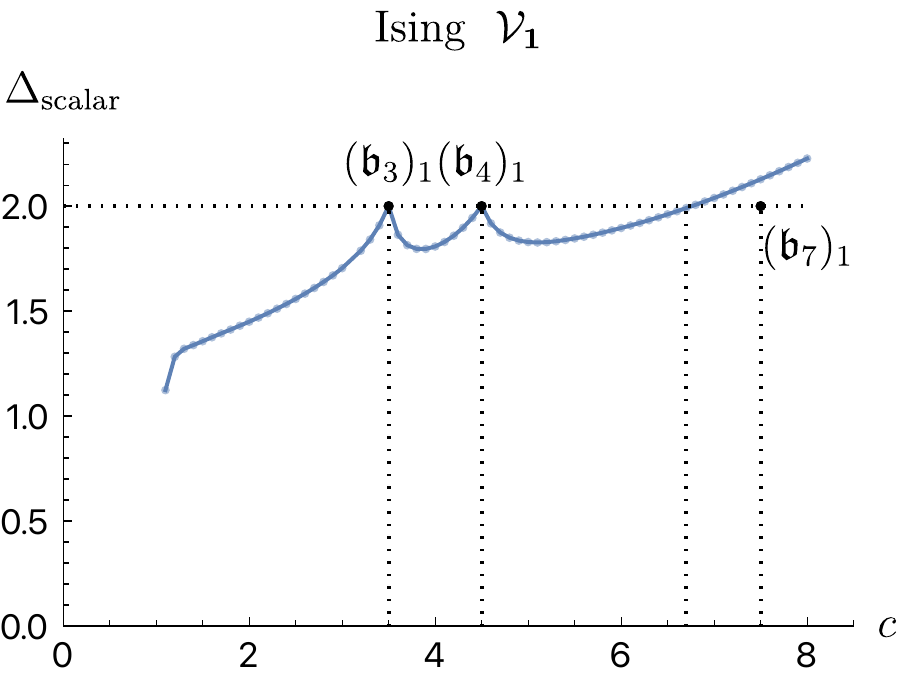}
    \includegraphics[height=.37\textwidth]{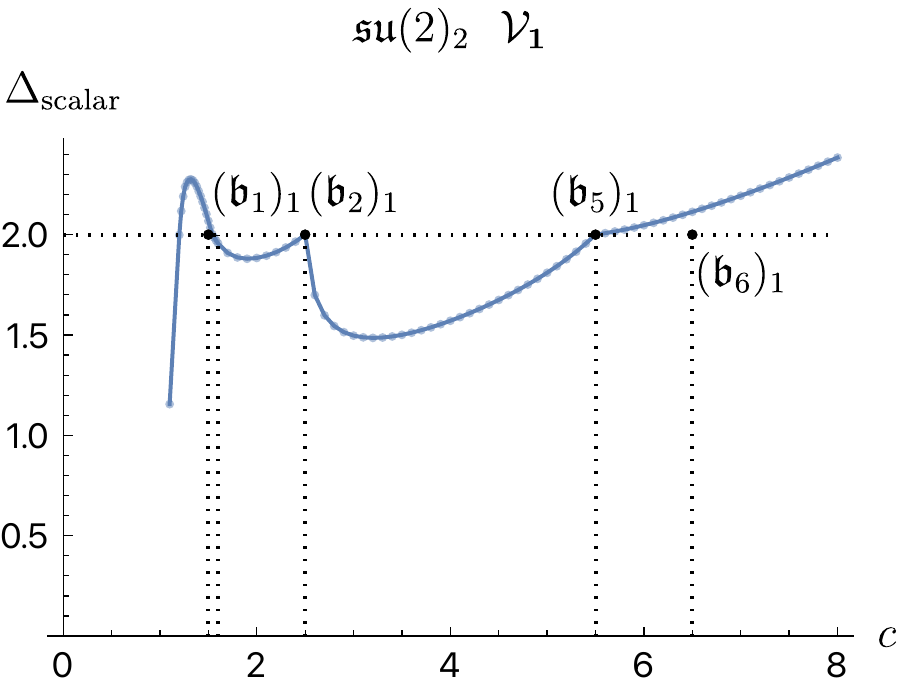}
    }
    \\
    ~
    \caption{Upper bounds on the scaling dimension of the lightest symmetry-preserving scalar local operator in any CFT with Fibonacci, Ising, and $\mathfrak{su}(2)_2$ categories, across a range of values for the central charge, $1 < c < 8$. This corresponds to the scalar subsector $(h=\bar h= \Delta/2)$ of $\cV_ \mathbf{1}$.
    }
    \label{Fig:Bounds-sc}
\end{figure}

For concreteness, we study the three categories reviewed in Section~\ref{Sec:Examples}, Fibonacci, Ising, and $\mathfrak{su}(2)_2$.  Since these categories can all be equipped with a modular braiding structure, the UMTC of the 2+1d TQFT is the product $\ZC \cong \cC \boxtimes \bar\cC$, as explained near the end of the Section \ref{sec:anyon}, and the modular data are simply the tensor products of the constituents.  The bounds on $\Delta$ are presented in Figure~\ref{Fig:Bounds-sc}. 
We find that $\cC$-protected gapless phases do not exist for the following ranges of the central charge:
\ie \label{crange}
    \text{Fibonacci}: & \quad 1 < c < 6.0,
    \\
    \text{Ising}: & \quad 1 < c < 6.7,
    \\
    \mathfrak{su}(2)_2: & \quad 1.6 < c < 5.5.
\fe 
In particular, the bounds are saturated at $\Delta = 2$ by the current bilinears in the $(\mathfrak{b}_n)_1$, $(\mathfrak{g}_2)_1$, and $(\mathfrak{f}_4)_1$ WZW models discussed in Section \ref{Sec:Examples}; see \eqref{Hc} and the discussion below.

\subsection{Bounds on Symmetry-Preserving/Violating Operators} \label{sec:violate}

\begin{figure}[h!]
    \centering
    \subfloat{
    \includegraphics[height=.37\textwidth]{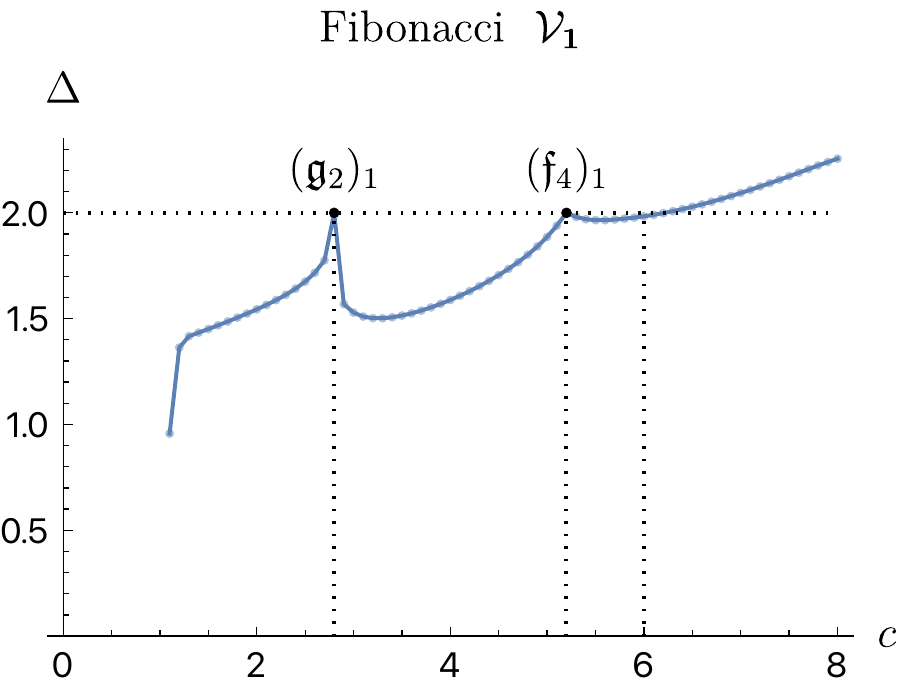}
    \includegraphics[height=.37\textwidth]{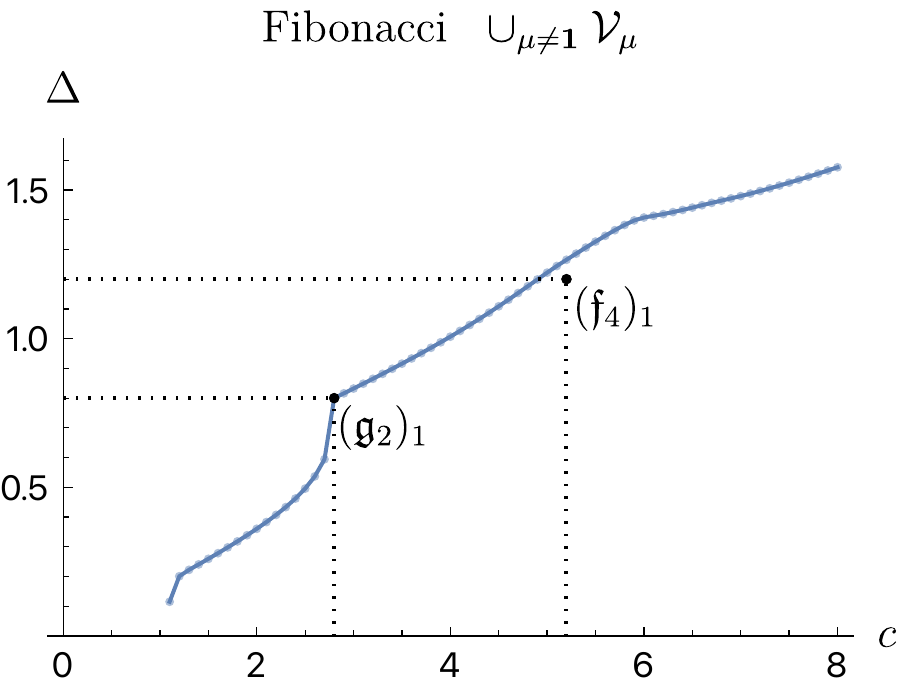}
    }
    \\
    ~
    \\
    \subfloat{
    \includegraphics[height=.37\textwidth]{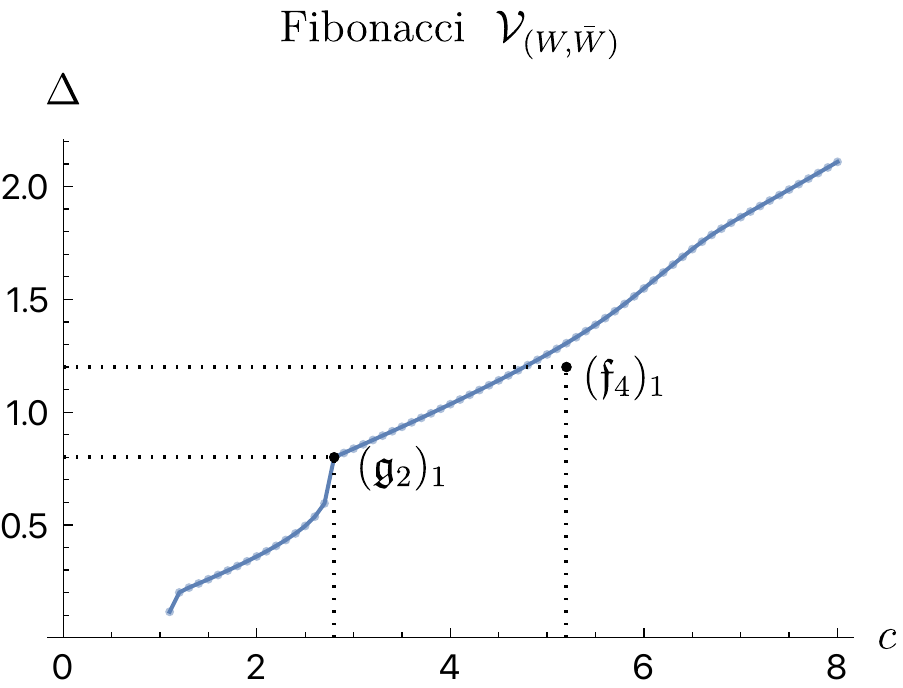}
    }
    \\
    ~
    \caption{Upper bounds on the lightest primary (of any Lorentz spin $h-\bar h$) in various anyon sectors in the case of the Fibonacci category, across a range of values for the central charge, $1 < c < 8$.  
    In particular, the bound in the $\cV_{(W,\bar W)}$ sector implies a bound on the lightest symmetry-violating local operator in any 1+1d CFT with a Fibonacci category symmetry.
    }
    \label{Fig:Fibo}
\end{figure}

\begin{figure}[h!]
    \centering
    \subfloat{
    \includegraphics[height=.37\textwidth]{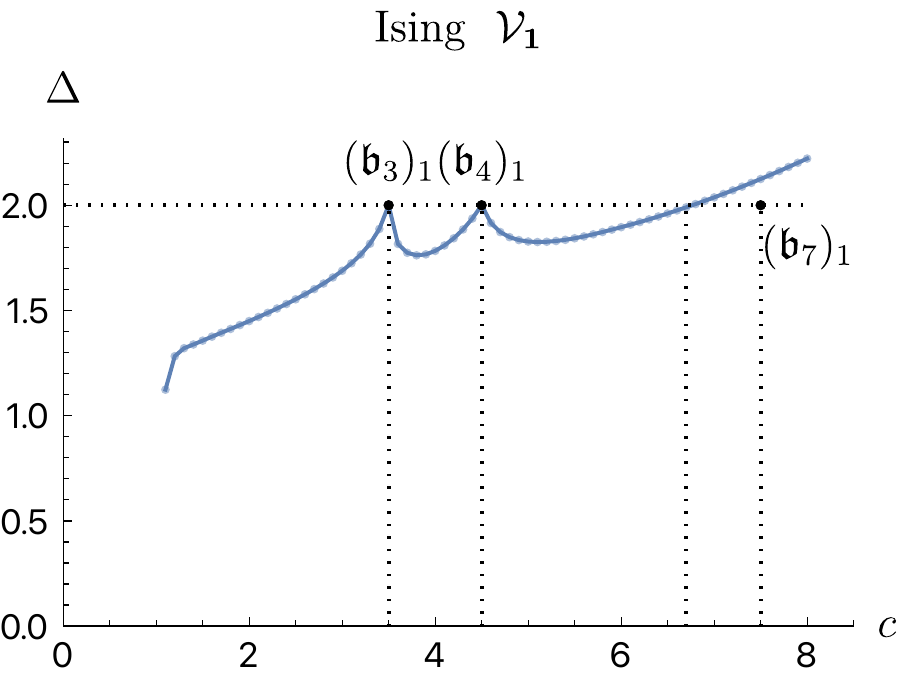}
    \includegraphics[height=.37\textwidth]{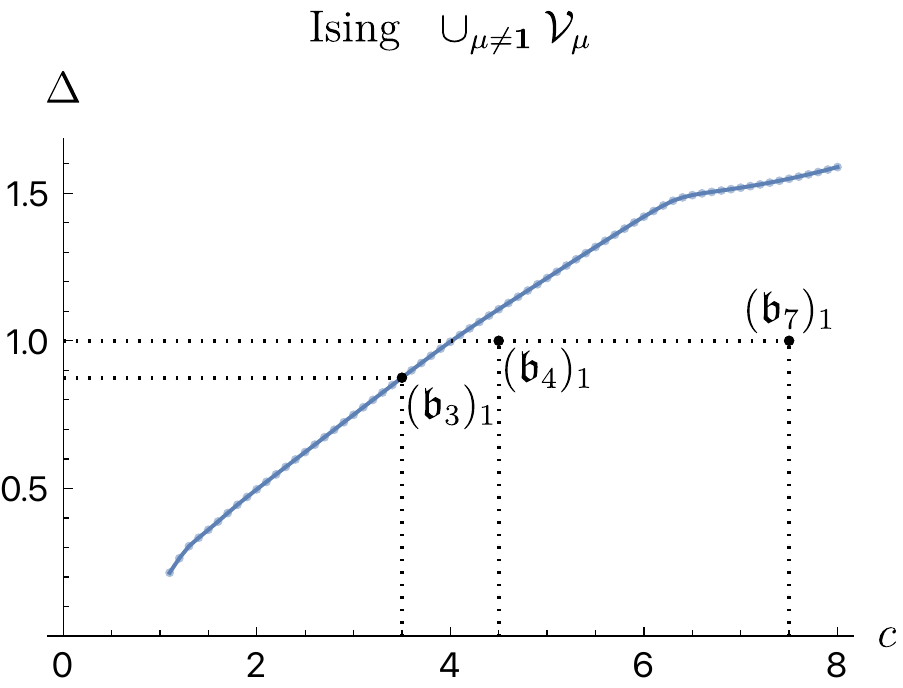}
    }
    \\
    ~
    \\
    \subfloat{
    \includegraphics[height=.37\textwidth]{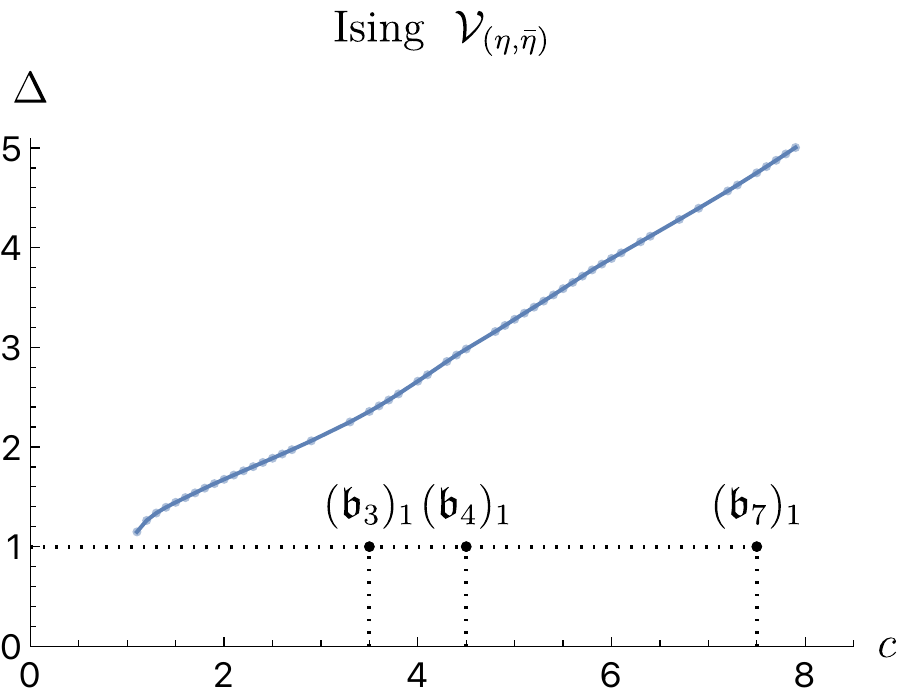}
    \includegraphics[height=.37\textwidth]{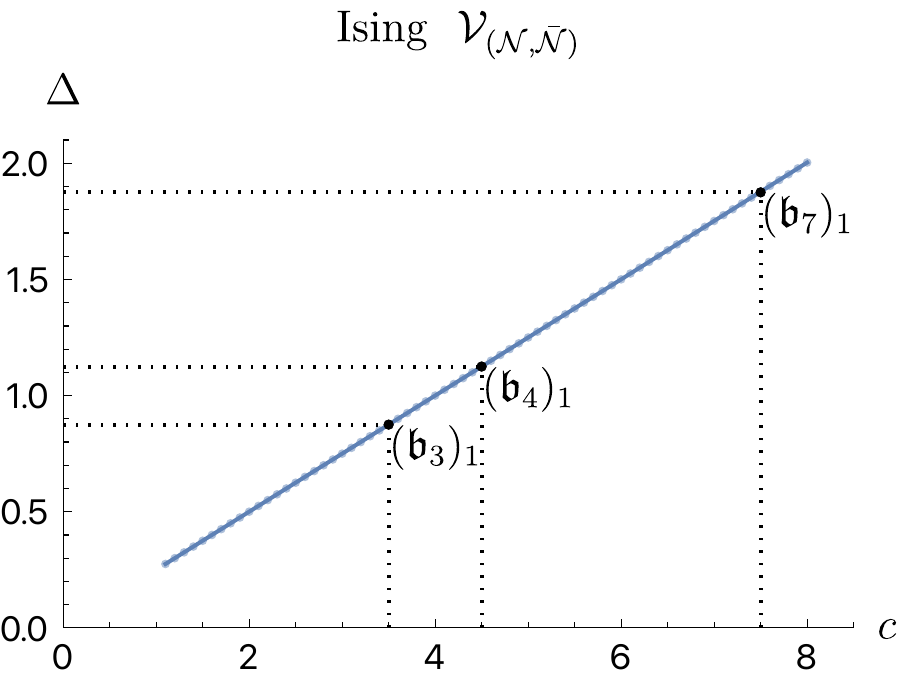}
    }
    \\
    ~
    \caption{Upper bounds on the lightest primary (of any Lorentz spin $h-\bar h$) in various anyon sectors in the case of the Ising category, across a range of values for the central charge, $1 < c < 8$.  In particular, the bounds in the $\cV_{(\eta,\bar \eta)}$ and $\cV_{(\cN,\bar \cN)}$ sectors imply bounds on two different kinds of symmetry-violating local operators in   1+1d. See the main text for detail. 
    }
    \label{Fig:Ising}
\end{figure}

\begin{figure}
    \centering
    \subfloat{
    \includegraphics[height=.37\textwidth]{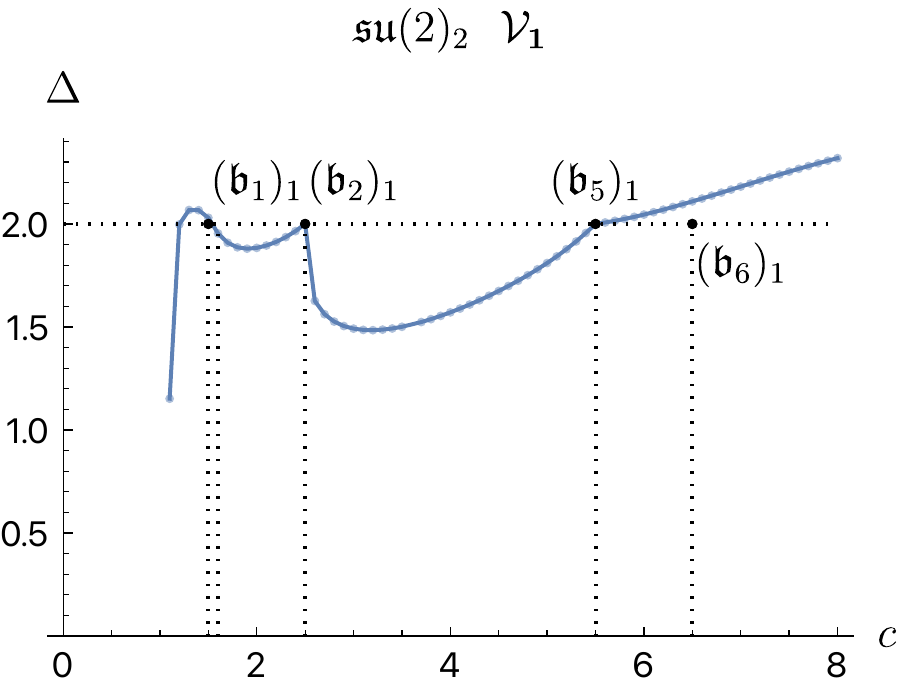}
    \includegraphics[height=.37\textwidth]{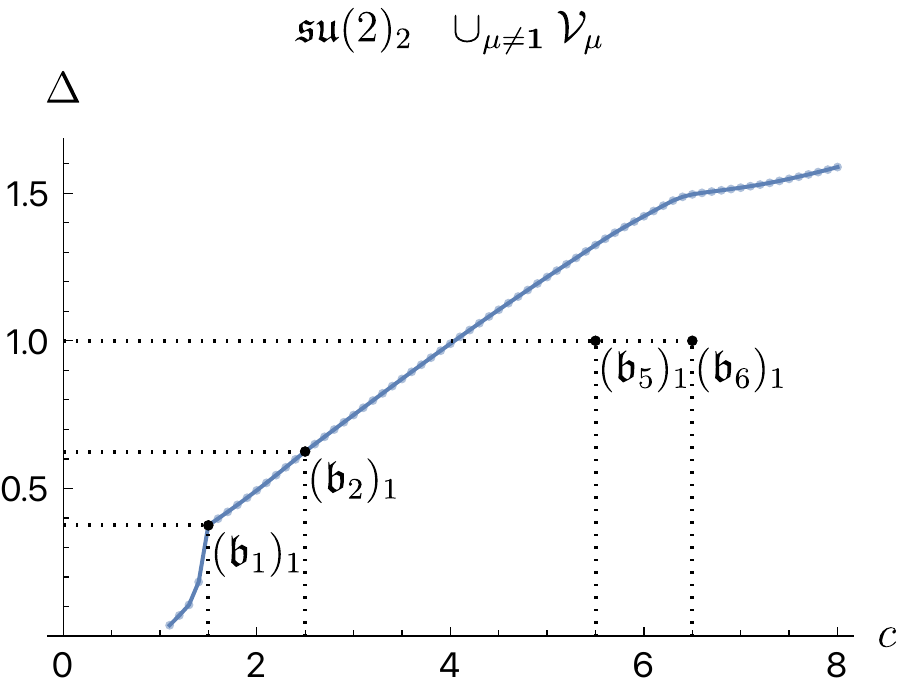}
    }
    \\
    ~
    \\
    \subfloat{
    \includegraphics[height=.37\textwidth]{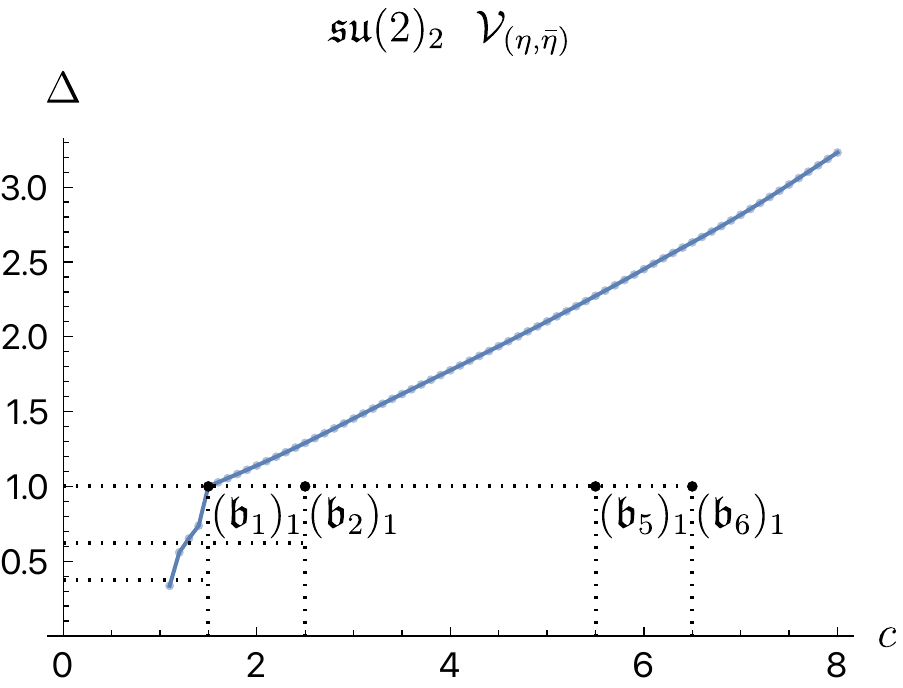}
    \includegraphics[height=.37\textwidth]{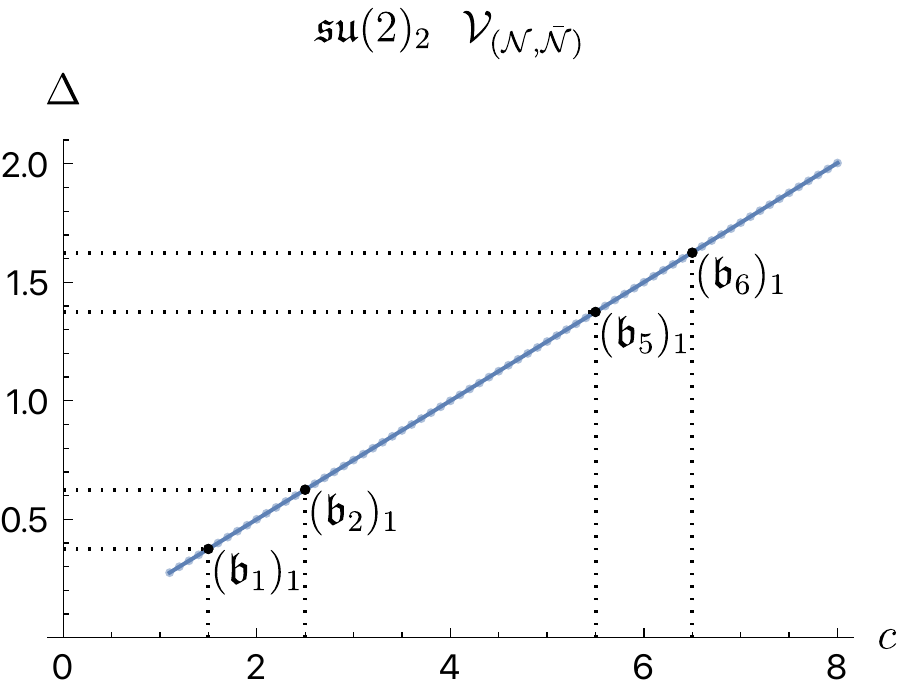}
    }
    \\
    ~
    \caption{Upper bounds on the lightest primary (of any Lorentz spin) in various anyon sectors in the case of the  $\mathfrak{su}(2)_2$ category, across a range of values for the central charge, $1 < c < 8$. 
    In particular, the bounds in the $\cV_{(\eta,\bar \eta)}$ and $\cV_{(\cN,\bar \cN)}$ sectors imply bounds on two different kinds of symmetry-violating local operators in 1+1d. See the main text for detail.
    }
    \label{Fig:su22}
\end{figure}

Next, we consider bounds on the gap in general anyon sectors $\cV_\mu$, not restricting to the scalar subsector.  
(See Appendix \ref{app:existence} for when a bound is expected to exist in a given sector.)
To be precise, we consider the gap in the spectra of non-degenerate Virasoro primaries, which in particular exclude conserved currents with weights $(h, 0)$ or $(0, \bar h)$ for $h, \bar h \in \bR$, in both the untwisted and twisted Hilbert spaces. 
The rest proceeds as described in the previous subsection.  
We present these bounds in Figures~\ref{Fig:Fibo}, \ref{Fig:Ising}, \ref{Fig:su22}. 
Using \eqref{HV}, these bounds can be translated into bounds on twisted Hilbert spaces of the original 1+1d CFT.

For the Fibonacci category, we present the bounds in Figure~\ref{Fig:Fibo} for three different sectors. Below we use \eqref{HV} to interpret these results as bounds on any 1+1d CFT with a Fibonacci category symmetry:
\begin{itemize}
\item  $\cV_\mathbf{1}$: This is an upper bound on the scaling dimension of the lightest symmetry-preserving primary operator (not necessarily a scalar) of the untwisted Hilbert space $\cH$ in 1+1d. 
It is similar to the bounds in Figure \ref{Fig:Bounds-sc}, but here we do not restrict to the scalar $h=\bar h$ subsector. 
\item $\cup_{\mu\neq \mathbf{1}}\cV_\mu$: This gives an upper bound on the union of the symmetry-violating sector of ${\cal H}$ and the twisted Hilbert space ${\cal H}_W$.
\item $\cV_{(W,\bar W)}$: Using \eqref{HV}, we can express the 1+1d untwisted Hilbert space  in terms of the anyon sectors, ${\cal H} =  \cV_\mathbf{1} \oplus\cV_{(W,\bar W)}$. Therefore, this bound in $\cV_{(W,\bar W)}$ can be understood as an upper bound on the lightest symmetry-violating local operator in 1+1d. 
\end{itemize}

For the Ising and $\mathfrak{su}(2)_2$ categories, we present the bounds in Figures \ref{Fig:Ising}, \ref{Fig:su22} for four different sectors. The first two bounds in  $\cV_\mathbf{1}$ and $\cup_{\mu\neq\mathbf{1}}\cV_\mu$ admit a similar 1+1d interpretation as in   the Fibonacci case above. For the other two:
\begin{itemize}
\item $\cV_{(\eta,\bar \eta)}$: This corresponds to states in the 1+1d untwisted Hilbert space $\cal H$ 
that preserve the $\bZ_2$ symmetry $\eta$ but violate  the 
non-invertible duality line $\cN$. 
More precisely, such a state $|\varepsilon\rangle$ obeys $\widehat\eta |\varepsilon \rangle = |\varepsilon\rangle\,,~\widehat \cN|\varepsilon\rangle =- \sqrt{2} |\varepsilon\rangle$. For instance, the local operator $\varepsilon_{\frac 12,\frac12}$ of the Ising CFT belongs to this sector. 
\item $\cV_{(\cN,\bar \cN)}$: This corresponds to the states in the 1+1d untwisted Hilbert space $\cal H$ that violate both $\eta$ and $\cN$.
More precisely, such a state $|\sigma\rangle$ obeys $\widehat\eta |\sigma \rangle =- |\sigma\rangle\,,~\widehat \cN|\sigma\rangle =0$. For instance, the local order/spin operator $\sigma_{\frac {1}{16},\frac{1}{16}}$ of the Ising CFT belongs to this sector. 
\end{itemize}

For the $\cV_{ (\cN, \bar\cN)}$ sector of either Ising or $\mathfrak{su}(2)_2$, the bound sits tantalizingly close to the simple formula $\frac{c}{4}$, which for $c = n+\frac12$ is realized by the $h=\bar h = {2n+1\over 16}$ state of the $(\mathfrak{b}_n)_1$ WZW model (see \eqref{Hso}).  It would be interesting to see if this formula gives the exact bound, perhaps by constructing an analytic functional \cite{Mazac:2016qev}.

The Ising and $\mathfrak{su}(2)_2$ categories of a non-spin CFT are related to the anomaly of an invertible $\bZ_2$ symmetry (together with $(-1)^F = \bZ_2^f$) in a spin CFT by fermionization \cite{Thorngren:2018bhj,Ji:2019ugf}. 
It would be interesting to compare our bounds with the fermionic modular bootstrap \cite{Grigoletto:2021zyv}.\footnote{For instance, the scalar bound in $\cV_\mathbf{1}$ for the Ising category, presented in Figure~\ref{Fig:Bounds-sc}, is qualitatively similar to the bound on the $\mathbb{Z}_2 \times \mathbb{Z}_2^f$-symmetric scalar states with $\nu=1$ in the untwisted NS Hilbert space in the fermionic modular bootstrap, presented in Figure~15 of \cite{Grigoletto:2021zyv}.  Here $\nu$ mod 8 labels the fermionic anomaly.
}

As explained above, the bounds on $\cV_{(a,\bar a)}$ with $a\neq 1$ correspond to bounds on the $\cC$-violating local operators of any 1+1d CFT with a fusion category symmetry $\cC$ (that can be lifted to an UMTC). 
This is analogous to the bound on the lightest $\bZ_2$-odd local operator obtained in \cite{Lin:2019kpn}. 
Importantly, in \cite{Lin:2019kpn}, it was found that such a bound only exists if the $\bZ_2$ is anomalous. 
This is because, for a non-anomalous $\bZ_2$ symmetry, there is a $\bZ_2$-symmetric trivially gapped phase with a unique vacuum, which presents a solution to the bootstrap equation such that there is no $\bZ_2$-odd local operator. 
For non-invertible symmetries, we expect a bound on the symmetry-violating sector to exist only for ``anomalous" fusion categories, in the sense that they are not compatible with a symmetric trivially gapped phase \cite{Chang:2018iay,Kaidi:2023maf}. 
(Mathematically, it means that they do not admit a fiber functor \cite{Thorngren:2019iar}.) 
All the fusion categories considered in the current paper are ``anomalous" in this sense.

\newpage

\section{Concluding Remarks}

In this paper, we presented universal bootstrap bounds on the scaling dimensions of states in different twisted Hilbert spaces of general 1+1d CFTs with Fibonacci, Ising, and $\mathfrak{su}(2)_2$ fusion category symmetries.  
All three fusion categories are intrinsically non-invertible \cite{Kaidi:2022uux,Kaidi:2022cpf,Kaidi:2023maf} in the sense that their Drinfeld centers are not Dijkgraaf-Witten gauge theories based on a finite group.
Some of these bounds are saturated by known RCFTs, including the $(\mathfrak{b}_n)_1, (\mathfrak{g}_2)_1, (\mathfrak{f}_4)_1$ WZW models. 
Our bootstrap method made use of the slab construction by coupling the CFT to a TQFT in one dimension higher. 

We were particularly interested in constraining the space of CFT that is stable under perturbations preserving a non-invertible global symmetry $\cC$. 
We rigorously proved that no such  $\cC$-protected gapless phase exists for certain ranges of the central charge $c$ presented in \eqref{crange}. 
Such CFTs commonly arise in the gapless phase of microscopic lattice models such as the anyonic chain, and our results place universal constraints on their phase diagrams.

Our modular bootstrap equations can be applied to more general systems. 
The most obvious generalization is to bootstrap CFTs with more general fusion category symmetries $\cC$, such as the Haagerup fusion category. 
It should be noted that all the categories considered in this paper have generalized anomalies in the sense that they are not compatible with a trivially gapped phase. 
(This is related to the existence  of a bootstrap bound on the symmetry-violating local operators as presented in Section \ref{sec:violate}.)
If we apply our bootstrap program to fusion categories that are non-anomalous, such as $\text{Rep}(G)$ with $G$ a finite non-abelian group or some of the Tambara-Yamagami categories   that admit a fiber functor, then there would be no bound on the symmetry-violating sector because of the existence of a fiber functor, that is, a trivially gapped symmetric phase.

Another exciting generalization is to replace the Turaev-Viro TQFT discussed in Section~\ref{sec:modular} with a general 2+1d TQFT that does not necessarily admit a gapped boundary condition (such as a TQFT with a nontrivial chiral central charge). The resulting bootstrap bound then places universal constraints on the edge modes of a 2+1d topological order, such as the fractional quantum Hall state.  We leave this for future investigation.

\section*{Acknowledgement}

We thank T.\ C.\ Huang, R.\ Lanzetta, S.\ Pal, and S.\ Seifnashri for useful comments on a draft. 
YL is supported by the Simons Collaboration Grant on the Non-Perturbative Bootstrap.
The work of SHS was supported in part by NSF grant PHY-2210182.
YL thanks the hospitality of the Kavli Institute of Theoretical Physics and the workshop ``Bootstrapping Quantum Gravity'' during the completion of the draft.
SHS thanks Harvard University for its hospitality during the course of this work. 
The authors of this paper were ordered alphabetically.

\appendix
 
\section{On the Existence of Bounds}\label{app:existence}

In this appendix, we explain when a bootstrap bound on the gap in a given sector exists.

In \cite{Lin:2019kpn}, it was found that for a non-anomalous $\bZ_2$ symmetry, a bound on the gap does not exist in the $\bZ_2$-odd sector of the untwisted Hilbert space, while such a bound does exist in the anomalous $\bZ_2$ case. 
The former fact is physically clear:  the lightest state charged under the $\bZ_2$ winding symmetry in a compact free boson CFT grows with the radius $R$ and can be arbitrarily large. 
 Furthermore, for a non-anomalous $\bZ_2$ symmetry, there is a $\bZ_2$-symmetric trivially gapped phase with a unique vacuum, which presents a solution to the bootstrap equation without any $\bZ_2$-odd operator (i.e.\ the $\bZ_2$ is not faithful). 
Similarly, for a generalized global symmetry, there cannot be a bound in the symmetry-violating sector if there exists a symmetric trivially gapped phase, in which case one can say that the symmetry is free of an anomaly in a generalized sense.

 Below we discuss in more detail a criterion for a bound to exist in a given sector(s) using the bootstrap equations.\footnote{In some cases, although the modular bootstrap equations alone do not yield a bound, the combined bootstrap system including other bootstrap equations, such as those for the four-point functions, give additional constraints. See \cite{Lanzetta:2022lze} for an example.} 
For a general symmetry, a modular bootstrap bound does not exist in a certain sector (or union of sectors) $\cV$ if the modular $S$- and $T$-matrices have a nontrivial, non-negative simultaneous eigenvector with eigenvalue $+1$ and vanishing component(s) in $\cV$. 
Such an eigenvector gives a solution to the modular bootstrap equation \eqref{Modular}  with empty $\cV$, and therefore there cannot be an upper bound in $\cV$.
  When the components of this eigenvector are non-negative integers, we can often understand the obstruction as coming from the existence of a 1+1d topological field theory realizing the symmetry, generalizing the non-faithful $\bZ_2$ interpretation above. 

In the non-anomalous $\bZ_2$ example, the modular matrices of the Drinfeld center---the toric code UMTC---is given in \eqref{TCMD}.  The simultaneous eigenspace with eigenvalue one is spanned by (in the $\{\mathbf{1}, e, m, f\}$ basis)
\ie 
    (1,1,0,0), \quad (1,0,1,0).
\fe 
telling us that a bound does not exist in either ${\cal V}_e$ (corresponding to the untwisted, $\bZ_2$ odd sector ${\cal H}^-$) or ${\cal V}_m$  (corresponding to the twisted, $\bZ_2$ even sector ${\cal H}_\eta^+$).  However, there is no nontrivial eigenvector with both components vanishing, and therefore an ``order-disorder'' bound is expected to exist in the union ${\cal V}_e \cup {\cal V}_m$, and was indeed established in \cite{Lin:2019kpn}.

As another example, consider an anomalous $\bZ_2$ symmetry. Its Drinfeld center is the $U(1)_2\times U(1)_{-2}$ UMTC, which is the low energy limit of the double semion model. Its modular matrices are given by   (in the $\{1,\eta\}\otimes  \{1,\bar \eta)\}$ basis)
\ie
S^\text{DS} = \frac12
\begin{pmatrix}
    1 & 1 & 1 & 1 \\
    1 & -1 & 1 & -1 \\
    1 & 1 & -1 & -1 \\
    1 & -1 & -1 & 1
\end{pmatrix},
\quad
T^\text{DS} = \text{Diag}(1, i, -i, 1).
\fe 
The simultaneous eigenspace with eigenvalue one is spanned by
\ie 
    (1,0,0,1).
\fe
By our general rule, there is no bound in ${\cal V}_{(\eta,1)}\cup {\cal V}_{(1,\bar\eta)}$, which  corresponds to the twisted Hilbert space ${\cal H}_\eta$. 
By contrast,  a bound exists as long as either or both of  ${\cal V}_\mathbf{1}$ and ${\cal V}_{(\eta, \bar \eta)}$ are included. 
For instance, in \cite{Lin:2019kpn}, a bound in the untwisted, $\bZ_2$ odd sector ${\cal H}^- \cong {\cal V}_{(\eta, \bar\eta)}$ was derived.

For the Fibonacci category, the simultaneous eigenspace of $S$ and $T$ with eigenvalue one is spanned by (in the $\{1,W\}\otimes  \{1,\bar W)\}$ basis)
\ie  
    (1,0,0,1).
 \fe 
We reach the same conclusion about the existence of bounds as for an anomalous $\bZ_2$.

For either the Ising or $\mathfrak{su}(2)_2$ fusion categories, 
the simultaneous eigenspace of $S$ and $T$ with eigenvalue one is spanned by (in the $\{1, \eta, \cN\} \otimes \{1, \bar\eta, \bar\cN\}$ basis)
\ie  
    (1,0,0,0,1,0,0,0,1).
\fe 
We expect a bound to  exist if any of ${\cal V}_\mathbf{1}, {\cal V}_{(\eta,\bar \eta)},{\cal V}_{(\cN,\bar\cN)}$ is included.
 
Generally, for any nontrivial unitary modular $S$-matrix (which does not have to be a Drinfeld center), we expect a bound on the gap in the union of all ${\cal V}_{\mu \neq \mathbf{1}}$, because $(1, 0, \dotsc, 0)$ cannot be an eigenvector with eigenvalue one. Had it been, it would imply that $S_{\mathbf{1},\mathbf{1}} = 1$, i.e.\ that the total quantum dimension of the UMTC is 1, in which case $S$ is trivial.  In defining the gap, we choose to exclude conserved currents in both the untwisted and twisted Hilbert spaces.  In RCFT, such a gap is given by the lightest scalar in the untwisted Hilbert space, or in other words, twice the minimal non-vacuum weight.

Similarly, we always expect a modular bootstrap bound on the gap in the symmetry-preserving sector ${\cal V}_\mathbf{1}$.  To see this, let us assume the contrary, that the modular $S$-matrix has a nontrivial non-negative eigenvector $e_\mu \ge 0$ with eigenvalue one and zero first component $e_1 = 0$.  Then it means that $0 = \sum_\mu S_{\mathbf{1},\mu} e_\mu$, which is a contradiction as $S_{\mathbf{1},\mu} = d_\mu S_{\mathbf{1},\mathbf{1}}$ is strictly positive. Here $d_\mu$ is the quantum dimension of the anyon line $\mu$.

To summarize, for any UMTC, bootstrap bounds in the following two sectors are expected to exist:
\begin{enumerate}
    \item The trivial anyon sector ${\cal V}_\mathbf{1}$.
    \item The union of all  ${\cal V}_\mu$ with $\mu \neq \mathbf{1}$.
\end{enumerate}
For the three unitary fusion categories studied in this paper, the following additional single-anyon-sector bounds are also expected to exist:
\begin{itemize}
    \item Fibonacci: ~ $\cV_{(W, \bar W)}$.
    \item Ising/$\mathfrak{su}(2)_2$: ~    $\cV_{(\eta, \bar\eta)}$ or $\cV_{(\cN, \bar\cN)}$.
\end{itemize}
These single-anyon-sectors correspond to symmetry-violating local operators of the 1+1d CFT as discussed in Section \ref{sec:modular}.

\bibliographystyle{JHEP}
\bibliography{cat}

\end{document}